\documentclass[AMA,LATO1COL]{WileyNJD-v2}

\articletype{Regular Issue Type}%

\received{17 September 2018}
\revised{XX XX 2018}
\accepted{XX XX 2018}

\usepackage{amsmath,amssymb,amsfonts}
\usepackage{url}
\usepackage{graphicx}
\usepackage{textcomp}
\usepackage{xcolor}
\def\BibTeX{{\rm B\kern-.05em{\sc i\kern-.025em b}\kern-.08em
		T\kern-.1667em\lower.7ex\hbox{E}\kern-.125emX}}
\usepackage{amssymb}

\usepackage{amsmath}
\usepackage{algorithm}
\usepackage{lipsum}
\usepackage{times}
\usepackage{url}
\usepackage{multirow}
\usepackage{rotating}
\usepackage{algpseudocode}
\usepackage{algorithm}
\usepackage{rotating}
\usepackage{wrapfig}

\usepackage{subcaption} 
\usepackage{kantlipsum} 
\usepackage{commath}
\allowdisplaybreaks

\usepackage{color, colortbl}
\definecolor{Gray}{gray}{0.9}

\definecolor{LightCyan}{rgb}{0.88,1,1}
\algnewcommand\algorithmicinput{\textbf{INPUT:}}
\algnewcommand\INPUT{\item[\algorithmicinput]}
\algnewcommand\algorithmicoutput{\textbf{OUTPUT:}}
\algnewcommand\OUTPUT{\item[\algorithmicoutput]}

\algdef{SE}[DOWHILE]{Do}{doWhile}{\algorithmicdo}[1]{\algorithmicwhile\ #1}

\makeatletter
\newcounter{algorithmbis}
\renewcommand{\thealgorithmbis}{\arabic{algorithmbis}}
\def\algorithmbis{\@ifnextchar[{\@algorithmbisa}{\@algorithmbisb}}
\def\@algorithmbisa[#1]{%
	\refstepcounter{algorithmbis}
	\trivlist
	\leftmargin\z@
	\itemindent\z@
	\labelsep\z@
	\item[\parbox{0.49\textwidth}{%
		\hrule
		\noindent\strut\textbf{Algorithm \thealgorithmbis} #1
		\hrule
	}]\hfil\vskip+0em%
}
\def\@algorithmbisb{\@algorithmbisa[]}

\makeatother

\begin{document}
	\title{FeatureAnalytics: An approach to derive relevant attributes for analyzing Android Malware}
	
	\author[1]{Deepa K.}
	\author[2]{Radhamani G.}
	\author[1]{Vinod P.}
	\author[3]{Mohammad Shojafar}
	\author[4]{Neeraj Kumar}
	\author[2]{Mauro Conti}
	
	\authormark{Deepa K.\textsc{et al}}
	
	\address[1]{\orgdiv{Research and Development Centre}, \orgname{Bharathiar University}, \orgaddress{\state{Coimbatore}, \country{India}}\\\email{deepaharinair@gmail.com}}
	
	\address[2]{\orgdiv{Department of IT and Science}, \orgname{Dr.GRD College of Science}, \orgaddress{\state{Coimbatore}, \country{India}}\\\email{radhamani@grd.edu.in}}
	
	\address[3]{\orgdiv{Department of Mathematics}, \orgname{University of Padua}, \orgaddress{\state{Padua}, \country{Italy}}\\\email{surename@math.unipd.it}}

	\address[4]{\orgname{Thapar Institute of Engineering and Technology}, \orgaddress{\state{Punjab}, \country{India}}\\\email{neeraj.kumar@thapar.edu}}
	
	\presentaddress{Department of Mathematics, University of Padua, Padua, 35131, Italy}
	
	\abstract[Summary]{Ever increasing number of Android malware, has always been a concern for cybersecurity professionals. Even though plenty of anti-malware solutions exist, a rational and pragmatic approach for the same is rare and has to be inspected further. In this paper, we propose a novel two-set feature selection approach based on Rough Set and Statistical Test named as RSST to extract relevant system calls. To address the problem of higher dimensional attribute set, we derived suboptimal system call space by applying the proposed feature selection method to maximize the separability between malware and benign samples. Comprehensive experiments conducted on a dataset consisting of 3500 samples with 30 RSST derived essential system calls resulted in an accuracy of 99.9\%, Area Under Curve~(AUC) of 1.0, with 1\% False Positive Rate (FPR). However, other feature selectors~(Information Gain, CFsSubsetEval, ChiSquare, FreqSel and Symmetric Uncertainty) used in the domain of malware analysis resulted in the accuracy of 95.5\% with 8.5\% FPR. Besides, empirical analysis of RSST derived system calls outperform other attributes such as permissions, opcodes, API, methods, call graphs, Droidbox attributes and network traces.}
	
	\jnlcitation{\cname{%
			\author{Deepa K.}, 
			\author{Radhamani G.}, 
			\author{Vinod P.}, 
			\author{M. Shojafar}, 
			\author{N. Kumar}, and 
			\author{M. Conti}} (\cyear{2018}), 
		\ctitle{FeatureAnalytics: An approach to derive relevant attributes for analyzing Android Malware}, \cjournal{Concurrency and Computation: Practice and Experience}, \cvol{2018;00:1--26}.}
	
	\maketitle
	\section{Introduction}
	\label{introduction}
	
	In the past few years, Android has been widely adopted as the preferred operating system of smartphones, tablets, and even Internet of Things (IoT) devices. In particular, smartphones being portable with its extensive computing capabilities have gained widespread attention than personal computers. Reports from~\cite{statista} state that in 2017 the sale of Android-based smartphones have surpassed 1.32 billion. The smartphone industry is steadily increasing and estimated to touch 1.71 billion in 2020~\cite{statista}. Since December 2017~\cite{statista}, over 3.5 million new apps are being uploaded to Google Play Store. Unfortunately, security issues with Android system evolve due to the tremendous growth of third-party app stores, which hosts numerous malware applications. Recently, Sophos lab~\cite{sophos}, reports submission of 10 million Android samples by the end of December 2017, of which 77\% of applications are identified as malware. Notably, the popularity of Android phones and its ubiquitous nature also attracted the adversary to exfiltrate critical information from compromised devices. Moreover, these malicious apps are used for phone-tapping, steal sensitive information, geographic locations,  and send premium rate messages. Considering the above-mentioned circumstances, immediate attention is required for enforcing security of smart devices from malignant applications.
	
	\par There are broadly two approaches for malware detection: (a) static analysis and (b) dynamic analysis methods. Traditionally, static analysis is used to create malware signatures. The sequence of instructions, strings, bits or hashes may be used to express signature. Even though signature-based techniques can rapidly identify malicious applications, they can be easily evaded by code polymorphism or techniques involving source code transformation. Additionally, signature-based detector proves ineffective for detecting zero-day malware. Dynamic analysis is also known as a behavior-based approach. Here, the antimalware engine evaluates actions of an app to determine if the application demands unauthorized access to the sensitive resources.
	
	\par Moreover, several reinforcement solutions are proposed to study the malware codes and its behaviors to realize the threats~\cite{faruki2015android}. To be precise, machine learning-based approaches~(MLA) in malware detection have gained increased acceptance due its increase in detection accuracy~\cite{Amos2013},~\cite{Gardiner2016}.
	
	\par In this context, several questions arise: Is it possible to present an Android malware detection framework that categorizes various applications and distinguishes between the malicious and benign ones? How can we select significant features, either using static or dynamic analysis, to identify the malware apps? How can we construct an optimal feature vector to improve classification? The goal of this paper is to shed light on these issues.
	\par In this paper, we present and investigate malware detection by developing the feature set comprising of system calls. The characteristic of the feature space is its representative power of exhibiting the behavior of an application. To showcase the indented actions of a monitoring application, Android Monkey~\cite{monkey} is employed to supply random inputs~(in the form of swipes and clicks, etc.) to the sample. Each event triggers the invocation of a set of the characteristic system calls. The extracted set of attributes might have irrelevant calls that do not help in the process of identifying malicious applications. Hence, a two-step feature selection method is proposed. Initially, an optimal feature vector is derived by applying rough set feature selection approach~\cite{zhang2004rough, swiniarski2001rough}. Further, to boost the performance of classifiers, we further synthesize the previous attributes using statistical test, precisely the Large Population Test~\cite{Massey2006}, to generate the list of prominent malware and benign attributes. These features are subsequently utilized to develop classification model, using algorithms such as AdaBoostM1-J48~\cite{Yoav1999}, Random forest~\cite{breiman2001}, and Rotation forest~\cite{Kuncheva2007}. The main contributions of this study are summarized as follows:
	
	\begin{enumerate}
		\item We propose a two-step feature selection approach inspired by the rough set and statistical test~(named as RSST), capable of eliminating irrelevant attributes~(i.e., system calls) for improving classifier detection performance. 
		
		\item We perform an extensive analysis to investigate the optimal feature vector that depicts enhanced results. This is ascertained by varying the number of system calls in the feature space. The results of LPT-based feature set exhibited an accuracy of 99.8\% with 0.001 False Positive Rate~(FPR).
		
		\item We perform extensive analysis using different categories of features extracted with static and dynamic analysis. In particular, static features considered are permissions, opcodes, APIs, and, dynamic features include system calls, network trace, system call graphs and attributes extracted from Droidbox~\cite{Droidbox}. Moreover, we demonstrate that the performance of classification model created with system calls are better compared to others.
		
		\item We thoroughly evaluated the performance of our proposed feature selection approach with other attribute selectors traditionally utilized in the domain of malware detection. The experiments show that set of system calls derived by our method can separate malware and legitimate instances with 99.9\% accuracy compared to alternate techniques. 
	\end{enumerate}
	
	\par The rest of this paper is structured as follows. Section~\ref{sec:relatedwork} discusses the related work in Android malware detection. Section~\ref{proposedmethod} presents our methodology. In Section~\ref{experiments}, we discuss the experiments, while we also present different setting of experiments to obtain relevant calls. 
	Section~\ref{comparative} presents the static and dynamic analysis to validate the efficiency of our proposed system call feature set. In Section~\ref{sec:performanceconventional} detection performance of our proposed two-step feature selection model is compared with other attribute selectors. After that, Section~\ref{discussions} describes some essential understandings and achievements of our solution and related experiments. We holistically compared our method against some state-of-the-art ones in Section~\ref{sec:comparepreviouswork}. We list the limitations of our method in Section~\ref{limitandfuturework}.  
	Finally, the conclusion is given in Section~\ref{conclusion}.
	
	\section{Related work}\label{sec:relatedwork}
	In the following, we will discuss the solutions adopted for Android malware analysis. Particularly, we will present static (see Section~\ref{sec:static}), dynamic (see Section~\ref{sec:dynamic}) and hybrid solutions (see  Section~\ref{sec:hybrid}) categories. In the following, we will briefly discuss each category.
	
	\subsection{Static Analysis}\label{sec:static}
	
	\par The paper~\cite{arp2014drebin} proposed DREBIN developed with the features extracted from both manifest files and bytecode and embedded all of the attributes into a joint vector space to detect the malicious apps using support vector machines (SVM). DREBIN detects 94\% detection rate of the malware samples with an FPR of 1\%. However, it is quite ineffective at detecting new types of malware because of the enormous size of the feature set. 
	
	\par Authors in ~\cite{zhou2012dissecting} present a systematic characterization of Android malware based on method of installation, activation mechanism and malicious payload. 
	
	\par Another solution~\cite{glodek2013rapid} combines application permissions, broadcast receivers, the presence of embedded Android applications and native code. They adopt random decision trees that built rules comprising of two or three features respectively.  
	
	\par A probabilistic discriminative model based on regularized logistic regression for detecting malicious apps from decompiled source code was proposed by authors in~\cite{cen2015probabilistic}. API calls and permissions were used as features. Further, features selection methods i.e., Information gain and Chi-squared were utilized for determining significant attributes. A comprehensive analysis on datasets collected from different sources was performed. Finally, classifier performance was evaluated using the metric like precision, recall, and Area Under Curve~(AUC).
	
	\par Droid Detective~\cite{talha2015apk} uses a rule-based approach to classify apps as benign and malware.  Combination of permissions together with their frequencies is utilized to create a set of rules. Subsequently, relevant rules essentially discriminating malware and benign samples were extensively investigated.
	
	\par Wei et al.~\cite{Wei2018} proposed a framework for identifying malapps and trusted applications along with the categorization of benign samples into diverse groups. Feature such as requested/used permissions, filtered intents, code-related information, restricted/suspicious API calls and hardware related attributes was extracted from a large collection of applications. Support Vector Machine (SVM) was used to rank attributes. Ensembles of SVM, CRT, Naive Bayes, Random forest and K-NN was considered to label an app into respective classes. The decision of unseen samples was arrived using majority voting. 
	
	\subsection{Dynamic Analysis}\label{sec:dynamic}
	\par In~\cite{ham2013analysis}, authors considered vectorized representation of CPU utilization, network traffic, power and memory consumption by apps as features. Information gain feature selection filtered 10 prominent attributes of total 32 features. Malware detection models employing Random forest, Naive Bayes, Logistic Regression and Support Vector Machine~(SVM) is created. An F-Measure of 0.993 with 0.998 AUC obtained with Random forest demonstrates the suitability of ensemble classifier for developing malware classification model.
	
	\par Authors in~\cite{amamra2016generative} presented a detection framework using system calls having the possibility to be implemented in the resource-constrained environment. To address this issue they proposed filtering and abstraction process on 200 popular applications. During the filtering phase, irrelevant system calls are eliminated to describe the behaviour of the applications. Later, system calls with identical functions are consolidated. However, we argue that abstraction phase might lose some important information affecting the identification of malicious samples from legitimate instances. Further, the return types and parameters passed as an argument to the calls are distinct, thus mapping of multiple calls with few representative ones is not always feasible.
	
	\par Authors in \cite{kim2014linux} present an ML classification system employing 59 Linux based features characterizing memory, CPU, and Network from the Android OS to detect malicious applications. The analysis was carried out by eliminating a set of attributes to estimate the performance of classification model. Finally, 36 out of 59 features learned with SVM resulted in 98.85\% accuracy with 0.67\% False Positive Rate.
	
	\par In~\cite{narudin2016evaluation},  TCP packets during active communication between the infected system and attacker server were used to build feature set. \textit{ClassifierSubsetEval} feature selection method implemented in WEKA filtered six out of 11 attributes. The algorithms such as Bayes network, multi-layer perceptron, decision tree(J48), $K$-nearest neighbor and Random forest were considered for developing learning models. The experimental results indicated 99.99\% accuracy.
	
	\par A client-side application capable of executing on the device for detecting deviations of legitimate apps from their malicious counterpart was proposed in~\cite{chekina}. The detection system consisted of a machine learning model trained with network trace. The study demonstrated that applications were easily distinguishable by analyzing the traffic patterns. Thus, the behavior of applications could be modelled by analyzing network behavior . Verification of whether the application behavior is what it claimed to be could also be performed. 
	
	\par Andromaly a light-weight host-based framework for anomaly detection on Android smartphones was discussed in~\cite{shabtai2012andromaly}. Andromaly monitors various system metrics, such as CPU usage, volume of data transferred through network, number of active processes and battery usage. Then, Andromaly receives the feature vectors from main service, analyze them (i.e., using rule-based, knowledge-based classifiers or anomaly detectors) to perform threat assessment with Threat Weighting Unit~(TWU), which is eventually used in the detection process. 
	
	\par 
	In~\cite{TONG201722}, sequence of system calls with different depth was used as features. Initially, set of malicious and benign applications were executed in a smartphone, later call logs were processed out side the device. System call name was extracted and used to create malicious and legitimate patterns with respect to experimentally determined threshold value. Experiments conducted on 2000 malicious and bengin applications resulted in an accuracy exceeding 90\%.  
	
	\subsection{Hybrid Analysis}\label{sec:hybrid}
	In~\cite{su2012smartphone}, authors present two-side malicious apps defense scanner using ML technique modelled on Random forest classifier. Firstly, the scanner executes the samples in a sandbox environment, and system calls are collected. Secondly, the categorized applications which are labelled as malware or benign are rechecked by monitoring the network activity of each app. Finally, Wrongly labelled files were corrected if any application depicted suspicious network activity.  
	
	\par The detection of smartphone malware using the subset of system calls, the weighted sum of permissions and combination of permissions was addressed in~\cite{canfora2013classifier}.Experimental study reported statistical difference in \texttt{open}, \texttt{read}, \texttt{recv} and \texttt{write} system calls. They also claim that these system calls can be used for appropriately classifying malware and trusted applications. The overall precision of approximately 85\% was obtained by estimating values for different evaluation parameters. Besides, in ~\cite{rocha2013hybrid}, authors provide information flow control along with declassification policies on unannotated programs with support to runtime security labels. Such solution presents hybrid approaches which cover dynamic labels and execution constraints to handle legacy and untrusted and mobile codes.  
	\par On the other hand, authors in~\cite{feldman2014manilyzer} present Manilyzer exploiting information system that adopts $KNN$, SVM, and C.45 classification algorithms. Their results confirm that they have 90\% accuracy in the classification of an app corpus over 617 obtained applications. However, as static analysis, they fail to generalize the patterns of new malware specimens, where they discuss detection by capturing attributes from network packets.

	\par Authors in~\cite{lin2015three} present a three-phase detection and classification framework: Permission-Based Detection~(PBD), System--Call Based Detector~(SBD), and classification of malwares into their respective types. Experiments on the dataset comprising of 933 benign and 265 malware resulted in 97\% True Positive Rate (TPR) with 3\%v False Positive Rate (FPR)and 98\% accuracy.
	
	\par After reviewing the previously published papers, we conclude that the researchers concentrated on improving the outcome of classification by deriving attributes or using feature selection methods commonly used in machine learning domain. Different from prior work, we focused on developing novel feature selection method for improving results along with investigating robust attributes which can be used for developing effective malware identification models.
	
	\section{Proposed Methodology}
	\label{proposedmethod}
\begin{figure*}[htpb]
		\centering
		{\includegraphics[scale=0.65]{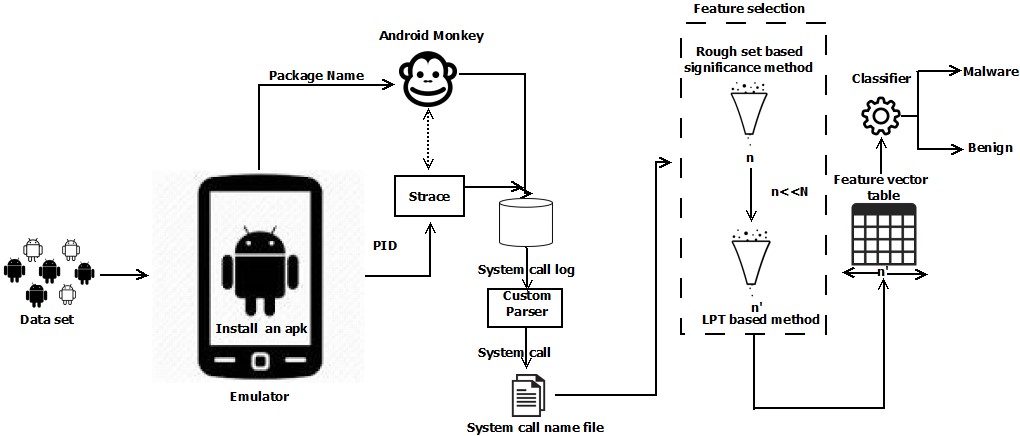}}
		\caption{Proposed framework, $n$ is the subset of reduced features obtained from attributes set of $N$ dimensions.}
		\label{fig:framework}
	\end{figure*}

	In this section, we present our approach for identifying malicious samples. The framework is shown in Fig.~\ref{fig:framework}. It is designed to contain multiple phases. In the initial stage, we collect malware and benign samples from multiple data stores. Each sample is installed in an emulator, which is subsequently interacted with Android Monkey. We use \texttt{strace} command to capture system call on the Android mobile application. System call logs are processed to filter call names, which are further used as features. In the subsequent phase, the feature set is refined using two-step feature selection approach. Initially, irrelevant calls are eliminated using rough set approach, later large population test is applied to determine discriminant system calls. Classification models are developed with the extracted attribute set, and finally, samples are separated into one of the two classes, i.e., either malware or benign. In the subsequent sections, we detail each phase involved in our proposed approach.

	\subsection{Dataset Description}
	The dataset consists of 2000 benign applications which are downloaded from Google Play Store~\cite{googleplaystore}, Chinese market~\cite{appchina}, Koodous~\cite{Koodous}, and third-party Android markets~\cite{1mobile}\cite{9apps}. Each sample is submitted to VirusTotal~\cite{virustotal}, an online antimalware service, to confirm applications are indeed legitimate. The malware dataset constitutes 1500 samples. A total of 554 malicious apps are randomly collected from Drebin project, 450 taken from Koodous-a collaborative platform for Android Malware analysis, consisting of analysis tools and combine social interactions.  Also,  a set of 496 ransomware apps~\cite{ransomware} is also used as a part of malware dataset.
	
	\subsection{Extraction of System calls}
	System call logs, during the execution of each sample, are considered as the feature for classifying a file as malware or benign. Linux system calls can be categorized based on operating system functionality such as process management, file management, memory management, device management, information management and communication models.  System calls act as an interface between user and kernel. All requests performed at the user mode are forwarded through a system call interface before its execution through the hardware. At any occasion, if a user intends to make a phone call using a call application, then the user request is transferred to Telephony Manager Service, to a set of library calls, which in turn results in multiple invocations of a system call. During the execution of a system call, control is transferred from user mode to kernel mode. When the system call is completed, then the control is returned back to the user mode. Thus, the interaction of a program with OS can be precisely exhibited by representing feature set with system calls or sequence of system calls.  Moreover, static features (e.g., permissions, metadata, opcodes, API's, and intents) are susceptible to change due to obfuscation, however, system calls are relatively resilient to obfuscation comparing attributes extracted during static analysis.
	
	\par Each application is installed in Android emulator using \texttt{adb install} command. Initially, we keep track of the \texttt{zygote} process which starts at \texttt{init}. Whenever a new application is launched, the \texttt{zygote} is forked. Using \texttt{strace} utility, we monitor the \texttt{zygote} and later filter out the process id, i.e., \texttt{pid} of the required application. Using Android Monkey, random events consisting of touches, clicks, and gestures, etc. are supplied to the application. In particular, we subject the application to SMS, phone and direction events one after another, to gather its actual behavior. During each event, system calls are recorded. Employing Android Monkey, 2500 events are subjected to an application, and the call trace is collected. The execution trace consists of system call name, parameters and return values. Using a customized parser, we extract call names which are utilized as features. Following steps are employed to extract system calls.
	
	\begin{itemize}
		\item [\textbf{S1:}]The .apk files are installed in emulator using \texttt{adb shell} command: \\\texttt{adb -s <emulator-id> install filename.apk}
		
		\item [\textbf{S2:}]Afterwards, the call trace are recorded with \texttt{strace} utility~\cite{strace}. The input to the \texttt{strace} is \texttt{pid} and the \texttt{logfilename}.\\
		\texttt{strace -p <pid> -o sdcard/logfilename}
		
		\item [\textbf{S3:}]  \texttt{aapt} command is run to obtain package name for an apk. The following command is used to interact with an application.\\\texttt{adb shell monkey -p <package-name> -v <\# events>}
		
		\item [\textbf{S4:}]  Other fake actions~(such as sending SMS, making/receiving calls or setting locations) are performed using some commands shown below:
		\begin{itemize}
			\item Connect to the emulator, using    
			telnet:\\\texttt{telnet localhost 5554}
			\item To make a phone call\\\texttt{gsm call<callerPhoneNumber>}
			\item Send an SMS\\\texttt{sms send <senderePhoneNumber> <textMessage>
			}
			\item To change geo-locations\\\texttt{geo fix <longtitude value> <latitude value>
			}
		\end{itemize}
		\item [\textbf{S5:}] The log file stored in emulator is copied to the device using \texttt{adb pull}.\\ 
		\texttt{adb -s emulator-<id> pull sdcard/logfilename destination-path}
		
		\item [\textbf{S6:}] The emulator instance is killed and the android device is restored to the previous clean state.
	\end{itemize}
	
	\subsection{Representation of Feature Vector}
	\label{feature}
	\par The goal of malware classification system is to map a collection of applications~(or \textit{apks}) into a fixed number of predefined categories i.e., malware~(M) or benign~(B). Hence, this is a supervised learning problem. To this end, the preliminary task is to transform each apk typically into a group of features. Formally speaking, each system call $s_i$ in this case corresponds to a feature. To adapt attributes into a feature vector, representative calls from samples are converted to a specific value. In conventional approach, feature vectors are represented as boolean value (presence/absence of an attribute is expressed as 1/0) or number of times $s_i$ occurs in the samples. Since, the classification is based on the contents~(i.e., system calls), an attribute weighting scheme known as \textit{Term Frequency-Inverse Document Frequency~(TF-IDF)}~\cite{Ramos2003} is utilized. Specifically, elements of each vector are the TF-IDF weight of a system call. This representation of feature vector assigns a higher weight to system calls that are typical of a sample, compared to calls that are relatively rare in the whole collection of instances. Thus, a collection of feature vectors are referred to us as Feature Vector Table~(FVT), which is a data structure consisting of $r$ rows of instances (vectors) and $N$ columns of system call (see Fig.~\ref{fig:fvt}). As supervised learning is used, each vector is labeled as malware ($M$) or benign ($B$).
	
	\begin{table}[h]
		\centering
		\caption{Representation of Feature Vector Table~(FVT). The elements of FVT~($tf$\hbox{-}$idf_{ij}$) designate the weights of system calls $j$ in the sample $i$; $r$ is the number of instances/samples; $N$ is the number of system calls; $x_j$ denotes $j$th malware/benign sample;  and, labels $M$ and $B$ denote Malicious and Benign samples.}
		\begin{tabular}{|c|c|c|c|c|c|c|}
			\hline
			\small
			\textbf{}&\multicolumn{5}{|c|}{System Calls}&\textbf{}\\\hline
			\textbf{} &\textbf{$s_1$} &\textbf{$s_2$} &\textbf{\ldots} &\textbf{$s_{(N-1)}$} &\textbf{$s_{N}$} &\textbf{Class} \\ \hline
			\textbf{$x_1$} &\textbf{$tf$\hbox{-}$idf_{11}$} &\textbf{$tf$\hbox{-}$idf_{12}$} &\textbf{\ldots} &\textbf{$tf$\hbox{-}$idf_{1(N-1)}$} &\textbf{$tf$\hbox{-}$idf_{1N}$} &\textbf{M} \\ \hline
			\textbf{$x_2$} &\textbf{$tf$\hbox{-}$idf_{21}$} &\textbf{$tf$\hbox{-}$idf_{22}$} &\textbf{\ldots} &\textbf{$tf$\hbox{-}$idf_{2(N-1)}$} &\textbf{$tf$\hbox{-}$idf_{2N}$} &\textbf{M} \\ \hline
			\textbf{$x_3$} &\textbf{$tf$\hbox{-}$idf_{31}$} &\textbf{$tf$\hbox{-}$idf_{32}$} &\textbf{\ldots} &\textbf{$tf$\hbox{-}$idf_{3(N-1)}$} &\textbf{$tf$\hbox{-}$idf_{3N}$} &\textbf{M} \\ \hline
			\textbf{\ldots} &\textbf{\ldots} &\textbf{\ldots} &\textbf{\ldots} &\textbf{\ldots} &\textbf{\ldots} &\textbf{\ldots} \\ \hline\textbf{\ldots} &\textbf{\ldots} &\textbf{\ldots} &\textbf{\ldots} &\textbf{\ldots} &\textbf{\ldots} &\textbf{\ldots} \\ \hline
			\textbf{$x_{k-2}$} &\textbf{$tf$\hbox{-}$idf_{(k-2)1}$} &\textbf{$tf$\hbox{-}$idf_{(k-2)2}$} &\textbf{\ldots} &\textbf{$tf$\hbox{-}$idf_{(k-2)(N-1)}$} &\textbf{$tf$\hbox{-}$idf_{(k-2)N}$} &\textbf{B} \\ \hline
			\textbf{$x_{k-1}$} &\textbf{$tf$\hbox{-}$idf_{(k-1)1}$} &\textbf{$tf$\hbox{-}$idf_{(k-1)2}$} &\textbf{\ldots} &\textbf{$tf$\hbox{-}$idf_{(k-1)(N-1)}$} &\textbf{$tf$\hbox{-}$idf_{(k-1)N}$} &\textbf{B} \\ \hline
			\textbf{$x_k$} &\textbf{$tf$\hbox{-}$idf_{k1}$} &\textbf{$tf$\hbox{-}$idf_{k2}$} &\textbf{\ldots} &\textbf{$tf$\hbox{-}$idf_{k(N-1)}$} &\textbf{$tf$\hbox{-}$idf_{kN}$} &\textbf{B} \\ \hline
			\textbf{\ldots} &\textbf{\ldots} &\textbf{\ldots} &\textbf{\ldots} &\textbf{\ldots} &\textbf{\ldots} &\textbf{\ldots} \\ \hline
			\textbf{$x_r$} &\textbf{$tf$\hbox{-}$idf_{r1}$} &\textbf{$tf$\hbox{-}$idf_{r2}$} &\textbf{\ldots} &\textbf{$tf$\hbox{-}$idf_{r(N-1)}$} &\textbf{$tf$\hbox{-}$idf_{rN}$} &\textbf{B} \\ \hline
			
		\end{tabular}
		\label{fig:fvt}
	\end{table}
	
	\subsection{Feature Selection Approach}
	\label{sec:featureselection}
	The property of an application that is being measured and characterizes it is known as feature~(or attribute). One of the dominant problems in machine learning over the past is identifying sub-optimal feature vector, having a strength equivalent to full feature space. Feature selection is a combinatorial optimization problem which aims to minimize redundant or irrelevant attributes. A feature is characterized as redundant if the information conveyed by the feature is more or less identical to the one or more features. On the contrary, a feature is considered as \textit{irrelevant}, if it does not carry essential information for identifying target classes.  
	\par Generally speaking, the objective of feature selection approach is to determine an optimal set from a finite set of a large number of attributes or reduce the number of possible solutions. Thus, in such problems, an exhaustive search is not feasible. Explicitly, in the context of malware detection, set of features (or attributes) are considered relevant if they can potentially identify target classes. Techniques using feature selection attempt to identify a small subset of attributes based on a fixed criterion. Relevance can be computed using specific statistical or information-theoretic approaches. Also, feature selection approach derives the useful attributes without changing its physical meaning. 
	
	\par In this work, we applied forward feature selection method. Feature selection was performed using search approach by applying SFE method. In this regard, the concepts of rough set theory~\cite{Hu1995} is used. Here, a table is represented as a tuple $T=(U,A)$, where $U$ represents the universe of instances set (i.e., apk files) and $A$ denote the set of attributes or objects, and $a \in A$ is an attribute instance. Let $P\subset A$ be the subset of features obtained by eliminating sparse features. Attribute set is partitioned into a set of conditional attributes $C$ (i.e., $M$ or $B$), and a set of decision attributes $D$ (i.e., system calls), respectively.
	
	\par Indiscernibility relation, $IND(P)$, is an equivalence relation defined as in equation~\eqref{eq:indp}.
	
	\begin{equation}
		\label{eq:indp}
		IND(P) = \{(x,y) \in U \times U, \forall a \in P, a(x)=a(y)\},
	\end{equation}
	where $a(x)$ is the feature value of object $x$; Here $a(x)=a(y)$ denotes that $x$ and $y$ are indiscernible with respect to $P$; and, $U/IND(P)$ represents all equivalence classes in $P$. 
	
	\noindent$L_\star$ is the lower approximation of $X$ which represents elements of $U$ that are surely in $X$. It denotes by
	\begin{equation}
		\label{eq:rlower}
		L_\star = U\{E \in U|IND(L):E\subset X\},
	\end{equation}
	$L^\star$ is the upper approximation of $X$ which represents the set that are possibly classified as elements in $X$. Equation \eqref{eq:rupper} contains the definition of upper approximation.
	\begin{equation}
		\label{eq:rupper}
		L^\star = U\{E \in U|IND(L):E\cap X \ne \phi\},
	\end{equation}
	\par The positive region is denoted as a set of applications of $U$ that can be classified with certainty to belong to classes $U/IND(D)$ using attribute $C$. In this paper, the significance of feature, i.e., system call, calculated from positive region $POS_C(D)$ is used as the criteria for feature selection. In order to construct a feature set, we estimated the reducts of conditional attributes (i.e., set of system calls) with respect to the decision attributes (i.e., the target classes). Johnson's greedy algorithm~\cite{jensen2007rough} can be used to determine reducts. Thus reducts eliminate all superfluous attributes from the feature set. Formally, the reduct of an conditional attribute $C$, w.r.t., decision features $D$ is the set of system calls $R\subseteq C$ that must be following properties (1) the classification metrics obtained with $R$ is similar to $C$, specifically the positive regions for $R$ and $C$ are identical, therefore, $POS_R(D)=POS_C(D)$ and (2) feature set of $R$ is minimal, thus $POS_{R-\{e\}}(D)\ne POS_R(D)$. It is represented using Equation~\eqref{eq:posregion}.
	\begin{equation}
		\label{eq:posregion}
		POS_C(D)=\bigcup_{X\in U/IND(D)}\underline{C}X,
	\end{equation}
	
	\par Finally, the significance of decision attributes~($\{M,B\} \in D$) on $C$ is defined as:
	\begin{equation}
		\Psi_C(D) = \frac{\mid POS_C(D)\mid}{\mid U\mid}.
		\label{eq:sig}
	\end{equation}
	where $\mid U \mid$ is the cardinality of a set $U$. A system call $s \in C$ is irrelevant in system call set $C$, if $\Psi_C(D) =\Psi_{C-\{s\}}(D)$, otherwise $s$ is regarded as relevant feature in $C$ with respect to target classes $(M/B)$. Therefore, set of attributes in reducts preserves the separation of classes. Subsequently, FVT records the TF-IDF score of system calls. These scores are mapped into four bins, where bin $B1$ is defined to contain TF-IDF values between 0.0-0.25, $B2$ contains values between 0.26-0.5, $B3$ contains values between 0.51-0.75, and $B4$ contains values between 0.76-1.0. Finally this representation of feature vector table is used to determine relevant attributes in the feature space.
	\begin{algorithm}
		\caption{System call with highest significant/dependency value.}
		\label{alg:significantsyscall}
		\normalsize
		\allowdisplaybreaks
		\begin{algorithmic}[1]
			\INPUT{U;A;D;V}
			\OUTPUT{S}
			\texttt{\textit{U}: All apk files\\
				\texttt{\textit{A}: Set of system calls}\\
				\texttt{\textit{D}: Set of decision attributes}\\
				\texttt{\textit{S}: Set of system call with highest significance/dependency value}\\
				\texttt{\textit{V[n][k]}: Feature occurrence matrix with $n$ samples \& $k$ system calls}}
			\For{$k$ in $A$}
			\For{$i \leftarrow 1$ to $\mid U \mid$}
			\Comment{$flag[i]$ denote a call included in a set $X$. The function match($i$,$j$) compares system calls $i$ and $j$ using binary search tree.}
			\State{flag[$i$] = 1;}
			\For{$j \leftarrow i+1$ to $\mid U \mid$}
			\If{(flag[$j$]==0) and \texttt{match(V[$i$][$k$],V[$j$][$k$])} and (D[$i$]==D[$j$])}
			\Statex\Comment{$x$ is a list of system calls with identical feature value and decision attribute}
			\State{X = X $\cup$ \{j\};}
			\State{flag[$j$] = 1;}
			\Comment{fromX is a flag, denotes a system call is appended to list $X$.}
			\State{fromX = 1;}
			\EndIf
			\EndFor
			\If{(flag[$j$]==0) and \texttt{match(V[$i$][$k$],V[$j$][$k$])} and (D[$i$]$\neq$D[$j$])}
			\State{Y = Y $\cup$ \{$j$\};}
			\State{fromY = 1;}
			\EndIf
			\EndFor                  
			\If{(fromX)}
			\Comment{$P$ is a dictionary of calls with identical values of feature $\&$ decision attribute.}
			\State{P[$k$][++$l$].append(X);}
			\State{fromX = 0;}
			\Else
			\Comment{$N$ is a dictionary of calls with identical feature values $\&$ different decision attribute.}
			\State{N[$k$][++$l$].append(Y);}
			\State{fromy = 0;}
			\EndIf
			
			\If{(flag[$i$-1] == 0)}
			\State{P[$k$][++$l$].append($i$);}
			\EndIf
			\Comment{Compute $\Psi_s(D)$ i.e., significance/dependency value for system call $k$.}
			\State{$\Psi[k] \leftarrow \frac{P[k]}{\mid U \mid}$} 
			\If{$\Psi[k] >$max$\Psi$}
			\State{max$\Psi$ $\leftarrow \Psi[k]$;}
			\State{$S \leftarrow S \cup k$;}
			\EndIf
			\For{$i\leftarrow 1$ to $\mid U \mid$}
			\State{flag[$i$] = 0;}
			\EndFor
			\EndFor \\
			\Return{S}
		\end{algorithmic}
	\end{algorithm}
	
	\begin{algorithm}
		\caption{Generate Reducts}
		\label{algo:generate_constraint}
		\allowdisplaybreaks
		\begin{algorithmic}[1]
			\INPUT{S;A;D}
			\OUTPUT{R}
			\Statex{\texttt{\textit{S}: a system call with highest significance/dependency value}}
			\Statex{\texttt{\textit{A}: be set of system calls}}
			\Statex{\texttt{\textit{D}: set of decision attributes}}
			\Statex{\texttt{\textit{R}: reduct}}
			\State{$R := S$;}
			\State{$temp := R$;}
			\Comment{$A\setminus\,R$ consists of all elements of $A$ which are not elements of $R$}
			\For{s $\in$ (A$\backslash$R)}
			\If{$\Psi_{R \cup s} > \Psi_{temp}$}
			\State{$temp := R \cup \{s\}$;}
			\EndIf   
			\State{$R := temp$;}
			\EndFor \\
			\Return{R}
		\end{algorithmic}
	\end{algorithm}
	
	\par The selection of a suitable/relevant subset of system calls are explained with steps listed in \textbf{Algorithm~\ref{alg:significantsyscall}} and\textbf{~\ref{algo:generate_constraint}}. Specifically, in Algorithm~\ref{alg:significantsyscall}, the line numbers 8 to 13 generate a set $X$ consisting of apks with identical values of conditional and decision attributes. Similarly, steps 14 to 18 is used to create a set $Y$ with apks having similar values of the conditional attributes but with dissimilar decision attributes. Steps 19-22 append the elements of sets $X$ and $Y$ to the dictionaries $P$ and $N$, respectively. The cardinality of set $P$ is subsequently used to determine the significance/dependency value of each system call $k$, illustrated in line number 29. The system call with highest significance/dependency value is determined~(steps 30 to 33) and returned to the procedure for generating reducts~(i.e., Algorithm~\ref{algo:generate_constraint}). In Algorithm~\ref{algo:generate_constraint}, the procedure for computing reducts takes \textit{three} parameters as input: a significant system call $S$, set of conditional attributes $A$, and decision attributes $D$. Algorithm~\ref{algo:generate_constraint} starts with the most significant system call. Subsequently, applies a forward approach to incrementally add attributes to reduct $R$ that has the highest significance value, refer line numbers 8-10. 
	
	\par \textbf{Complexity Analysis:} The time spent for computing reducts are related to the amount of comparison of feature vector, and all possible values of the vectors. In our case, the continuous values of elements are mapped to one of the possible bins~($\{B1, B2, B3, B4\}$). In our case, we have two classes and four bins, hence, we have 8 possible cases. Thus, maximum values of feature vector cannot be more than $8 \times \mid U\mid$, which itself is a huge number. Moreover, an in-accurate design of algorithm would require $O(C\cdot U^2)$ as the worst case time complexity. However, in our implementation, feature vectors are represented as a binary tree. In a nutshell, in our solution, the overall worst case time complexity for comparing feature vectors is estimated as $O(C\cdot\log U)\ll O(C\cdot U^2)$. 
	
	\subsection{An Illustrative Example}
	\label{sec:example}
	Let us consider an example to illustrate the selection of relevant system calls using the proposed feature selection approach. To demonstrate the procedure, we make use of an example feature vector table as shown in Table~\ref{tab:bins}. There are three system calls (i.e., $s_1$, $s_2$, $s_3$), seven applications (i.e., $x_1$-$x_7$) and a decision attribute $D$. The decision attribute consists of two values $M$ or $B$. In this example, each application $x_i$ is represented as a vector, the elements of a vector is mapped to bins, i.e., $\{B1,B2,B3,B4\}$ for a system call $s_j$~(see Table~\ref{tab:bins}).
	
	\begin{table}[h]
		\centering
		\caption{Mapping System calls to bins. Bin $B1$ ranges from 0.0 to 0.25, $B2$ is from 0.26-0.5, $B3$ lies between 0.56-0.75 and finally $B4$ is between 0.76-1.0.}
		\begin{tabular}{|c|c|c|c|c|}
			\hline
			\textbf{Samples} & $s_1$ & $s_2$& $s_3$ & \textbf{\textit{D}}\\ \hline
			$x_1$ & $B1$ & $B4$ & $B1$ & $B$\\ \hline
			$x_2$ & $B2$ & $B1$ & $B2$ & $M$\\ \hline
			$x_3$ & $B2$ & $B1$ & $B2$ & $M$\\ \hline
			$x_4$ & $B2$ & $B2$ & $B1$ & $B$\\ \hline
			$x_5$ & $B3$ & $B2$ & $B4$ & $M$\\ \hline
			$x_6$ & $B1$ & $B2$ & $B3$ & $B$\\ \hline
			$x_7$ & $B3$ & $B2$ & $B3$ & $M$\\ \hline
		\end{tabular}
		\label{tab:bins}
	\end{table}
	\par The procedure begins with an empty set of reduct $R$. Each system call is selected and its significance is computed, using Equation~\ref{eq:sig}. The call with highest significance value is selected and assigned to the reduct set $R$~(refer Algorithms~\ref{alg:significantsyscall} and \ref{algo:generate_constraint}). In this example, system call $s_3$ is added to $R$.
	\begin{equation*}
		U/A_D=\{ \underbrace{\{x_2,x_3,x_5,x_7}_{M}\},\underbrace{\{x_1,x_4,x_6\}}_{B}\},
	\end{equation*}
	\begin{equation*}
		U/A_{s_1}=\{\underbrace{\{x_1,x_6}_{B_1}\},\underbrace{\{x_3,x_4\}}_{B_2}, \underbrace{\{x_5,x_7\}}_{B_3}\},
	\end{equation*}
	\begin{equation*}
		\Psi_{s_1} = \frac{\mid \{x_5,x_7\}\mid}{\mid \{x_1,x_2,x_3,x_4,x_5,x_6,x_7\} \mid} = \frac{2}{7} = 0.28,
	\end{equation*}
	\begin{equation*}
		U/A_{s_2}=\{\underbrace{\{x_2,x_3}_{B_1}\},\underbrace{\{x_4,x_5,x_6,x_7\}}_{B_2}, \underbrace{\{x_1\}}_{B_4}\},
	\end{equation*}
	\begin{equation*}
		\Psi_{s_2} = \frac{\mid \{x_1,x_2,x_3\}\mid}{\mid \{x_1,x_2,x_3,x_4,x_5,x_6,x_7\} \mid} = \frac{3}{7} = 0.42,
	\end{equation*}
	\begin{equation*}
		U/A_{s_3}=\{\underbrace{\{x_1,x_4}_{B_1}\},\underbrace{\{x_2,x_3\}}_{B_2}, \underbrace{\{x_6, x_7\}}_{B_3}, \underbrace{\{x_5\}}_{B_4}\},
	\end{equation*}
	\begin{equation*}
		\Psi_{s_3} = \frac{\mid \{x_1,x_2,x_3,x_4,x_5\}\mid}{\mid \{x_1,x_2,x_3,x_4,x_5,x_6,x_7\} \mid} = \frac{5}{7} = 0.71,
	\end{equation*}
	\[\therefore R \leftarrow \{s_3\},\]
	
	\par Subsequently other attributes~($s_1$ or $s_2$) are added to $R$ by evaluating the significance of a system call with each attributes in $R$~(i.e, $s_3$). Hence, a forward feature selection strategy is employed.
	
	\begin{equation*}
		U/A_{\{s_1,s_3\}}=\{\{x_1\}\{x_2, x_3\},\{x_5\}, \{x_6\},\{x_7\}\},
	\end{equation*}
	\begin{equation*}
		\Psi_{\{s_1,s_3\}} = \frac{\mid \{x_1,x_2,x_3,x_4,x_5,x_6,x_7\}\mid}{\mid \{x_1,x_2,x_3,x_4,x_5,x_6,x_7\} \mid} = \frac{7}{7} = 1.0,
	\end{equation*}
	\begin{equation*}
		U/A_{\{s_2,s_3\}}=\{\{x_2,x_3\}\{x_1\},\{x_4\}, \{x_5\},\{x_6,x_7\}\},
	\end{equation*}
	\begin{equation*}
		\Psi_{\{s_2,s_3\}} = \frac{\mid \{x_1,x_2,x_3,x_4,x_5\}\mid}{\mid \{x_1,x_2,x_3,x_4,x_5,x_6,x_7\} \mid} = \frac{5}{7} = 0.714,
	\end{equation*}
	\[\therefore R \leftarrow \{s_1,s_3\}.\]
	Finally, the feature set $\{s_1,s_3\}$ is considered as feature set and eventually utilized to construct classification model.
	
	\subsection{Classification Phase}
	\label{classification}
	After the construction of feature set, our system creates classification models using three algorithms. In contrast to conventional signature-based scanners, the machine learning-based models require fewer updates, due to the fact that less number of malware is reported to form new families~\cite{avtest2016}. We employ commonly used classification algorithms reported in malware detection process. Classification algorithms such as Random forest~\cite{breiman2001}, Rotation forest~\cite{Kuncheva2007}, AdaBoost~(with J48 as base classifier) implemented in WEKA~\cite{WEKA} are considered.

	\subsection{Evaluation Parameters}
	\label{evalmetrics}
	To evaluate the effectiveness of our proposed method, we used classical evaluation metrics applied in machine learning;
	\begin{itemize}
		\item True Positive (TP): it indicates number of malicious applications that are appropriately identified.
		\item True Negative (TN): it denotes the number of accurately classified benign instances.
		\item False Positive (FP): it signifies the number of wrongly classified benign instances as malware applications.
		\item False Negative (FN): it indicates malware instances wrongly classified as legitimate application.
	\end{itemize}
	
	\par Using above mentioned criteria, following metrics are used to measure the effectiveness of our proposed system:
	
	\begin{itemize}
		\item \textbf{Accuracy (Acc):} Acc is the number of applications that the classifier correctly detects, divided by the number of malicious and legitimate applications.
		\begin{equation}
			Acc = \frac{TP+TN}{TP+FN+TN+FP}.		\end{equation}
		\item \textbf{False Positive Ratio (FPR):} FPR is the number of misclassified legitimate applications, divided by the number of benign applications.
		\begin{equation}
			FPR = \frac{FP}{FP+TN}.
		\end{equation}
		\item \textbf{Area Under Curve (AUC):} AUC is used to combine FPR and TPR together\cite{idrees2017pindroid}. In particular, AUC measures the tradeoff between TPR and FPR. Intrinsic goal of AUC is to solve situation where data set consists of imbalanced samples (or skewed sample distribution), and it is required that the model is not over-fitted to class consisting of higher number of instances. The value of AUC is between 0 and 1, AUC value 1 means the prediction is appropriate, it is reasonable if the value is greater than 0.5, however, if the value is less than 0.5, then we must reverse the decision of classification model.
	\end{itemize}
	\begin{equation}
		AUC = \frac{1}{2}\left(\frac{TP}{TP+FP}+\frac{TN}{TN+FP}\right).
	\end{equation}
	\section{Evaluation of Results}
	\label{experiments}
	The experiments are conducted on system with Intel core i7, 2 GHz quad-core processor and 8GB internal memory. We evaluate each classification model by a 10-fold cross-validation [40],[20] procedure to develop optimum model having improved generalization capability. Dataset is divided into ten equal subsets with 90\% of the set used for developing training model and remaining 10\% of instances used as test set.  Extensive analysis are performed on extracted features using static and dynamic mechanisms. Following sections present the experiments and analysis of the work.
	
	\subsection{Performance obtained with System call attributes}
	The effectiveness of classification system is evaluated under following settings:
	\begin{enumerate}
		\item Outcome of classification obtained with prominent benign system calls
		\item Performance of model developed using set of malicious system calls
		\item Classifier results ascertained with subset of feature space derived using statistical test. Furthermore, the analysis is conducted using significant benign and malicious system call set.
	\end{enumerate}
	
	\subsubsection{Evaluation on significant benign attributes}
	\label{sec:sigleg}
	The trusted application is executed in the emulator and applying the procedure discussed in Section~\ref{proposedmethod}; we extract system call names. The collection of call names are considered as attributes. Later, irrelevant attributes are removed using feature selection approach discussed in Section~\ref{sec:featureselection}. Using filtered system calls, classification models are generated, and the performance is evaluated with 10-fold cross-validation. We observe that Rotation forest and Random forest relatively yield similar performance. Table~\ref{tab:bensyscall} shows the weighted average of different metrics~(refer to the last row), both Rotation forest and Random forest resulted in AUC value 1.0, with an accuracy in the range of 99.42-99.54\%, and FPR of 0.005 and 0.01, respectively.
	
	\par In Table~\ref{tab:bensyscall}, we see that Random forest~\cite{breiman2001} with 80 system calls result in an AUC value of 1.0 with an FPR of 0.003. Model created with Rotation forest~\cite{Kuncheva2007} provides an AUC value of 1.0 using 50 system calls. However, the best outcome (i.e., FPR of 0.009 and AUC of 0.972) with AdaBoost is obtained with 80 attributes.
	
	\begin{table*}[!htpb]
		\centering
		\caption{Performance obtained with system calls extracted from benign samples. Accuracy and FPR are shown in percentage. The values of AUC are in range of 0-1.}
		{
			\begin{tabular}{|c|c|c|c|c|c|c|c|c|c|}
				\hline
				\multirow{1}{*}{\textbf{Feature}} & \multicolumn{3}{c|}{\textbf{Random forest}} & \multicolumn{3}{c|}{\textbf{Rotation forest}}& \multicolumn{3}{c|}{\textbf{AdaBoostM1}}\\\cline{2-10}
				\textbf{Length}& \textbf{Acc} & \textbf{FPR} & \textbf{AUC} & \textbf{Acc.} & \textbf{FPR} & \textbf{AUC} & \textbf{Acc} & \textbf{FPR} & \textbf{AUC}\\ \hline
				10 & 99.063 & 0.009 & 1.0 & 97.659 & 0.025 & 0.999 & 77.157 & 0.258 & 0.823\\ \hline
				20 & 99.156 & 0.004 & 1.0 & 98.313 & 0.018 & 1.0 & 89.316 & 0.108 & 0.962\\ \hline
				30 & 99.250 & 0.008 & 1.0 & 99.813 & 0.002 & 1.0 & 91.471 & 0.089 & 0.964\\ \hline
				40 & 99.625 & 0.004 & 1.0 & 99.813 & 0.002 & 1.0 & 90.253 & 0.099 & 0.969\\ \hline
				50 & 99.531 & 0.005 & 1.0 & 100 & 0 & 1.0 & 90.534 & 0.097 & 0.969\\ \hline
				60 & 99.531 & 0.005 & 1.0 & 99.906 & 0.001 & 1.0 & 89.784 & 0.106 & 0.968\\ \hline
				70 & 99.250 & 0.008 & 1.0 & 100 & 0 & 1.0 & 89.972 & 0.105 & 0.968\\ \hline
				80 & 99.719 & 0.003 &  1.0 &  99.906 &  0.001 & 1.0 &  90.347 & 0.101 &  0.972\\ \hline
				90 & 99.531 & 0.005 & 1.0 & 100 & 0 & 1.0 &  90.347  & 0.101 & 0.972\\ \hline
				92 & 99.531 & 0.005 & 1.0 & 100 & 0 & 1.0 & 90.347 & 0.101 & 0.972\\ \hline
				\textbf{Average}& \textbf{99.42} & \textbf{0.005} & \textbf{1.0} & \textbf{99.54} & \textbf{0.005} & \textbf{1.00} & \textbf{88.95} & \textbf{0.12} & \textbf{0.95}\\\hline
				\textbf{std-deviation}&\textbf{0.220}&\textbf{0.002}&\textbf{0}&\textbf{0.837}&\textbf{0.009}&\textbf{0}&\textbf{4.182}&\textbf{0.050}&\textbf{0.046}\\\hline
				
			\end{tabular}}
			\label{tab:bensyscall}
		\end{table*}
		
		\subsubsection{Performance on malware attributes}
		\label{sec:sigmal}
		We observe from Table~\ref{tab:malsyscall} that Random forest provided an AUC in range of 0.99-1.0 with FPR between 0.005-0.013 and accuracy in range of 99.344-99.787\%. The weighted average of evaluation metrics obtained for Random forest is better compared to Rotation forest and AdaBoostM1. Also, highest accuracy of 99.782\% with 10 system calls are obtained with Random forest, proving its efficacy for constructing malware detection model.
		
		\begin{table*}[!htbp]
			\centering
			\caption{Performance obtained with system calls that are extracted from malware samples.}
			{
				\begin{tabular}{|c|c|c|c|c|c|c|c|c|c|}
					\hline
					\multirow{1}{*}{\textbf{Feature}} & \multicolumn{3}{c|}{\textbf{Random Forest}} & \multicolumn{3}{c|}{\textbf{Rotation Forest}}& \multicolumn{3}{c|}{\textbf{AdaBoostM1}}\\\cline{2-10}
					\textbf{Length}& \textbf{Acc(\%)} & \textbf{FPR(\%)} & \textbf{AUC} & \textbf{Acc(\%)} & \textbf{FPR}(\%) & \textbf{AUC} & \textbf{Acc(\%)} & \textbf{FPR(\%)} & \textbf{AUC}\\ \hline
					10 &  99.782 & 0.013 & 0.999 & 94.845 & 0.05 & 0.988 & 74.157 & 0.167 & 0.898\\ \hline
					20 &  99.438 & 0.006 & 0.999 & 94.845 & 0.05 & 0.988 & 86.129 & 0.138 & 0.94\\ \hline
					30 &  99.156 & 0.009 & 0.999 & 99.25 & 0.008 & 0.999 & 86.129 & 0.138 & 0.94\\ \hline
					40 &  99.531 & 0.005 & 1.0 & 99.813 & 0.002 & 1.0 & 86.036 & 0.139 & 0.947\\ \hline
					50 & 99.531 & 0.005 & 1.0 & 99.719 & 0.003 & 1.0 & 87.816 & 0.139 & 0.949\\ \hline
					60 & 99.531 & 0.005 & 1.0 & 99.906 & 0.001 & 1.0 & 89.316 & 0.106 & 0.968\\ \hline
					70 &  99.438 & 0.006  & 1.0 & 99.906 & 0.001 & 1.0 & 90.815 & 0.097  & 0.973\\ \hline
					80 & 99.438  & 0.006  & 1.0 & 100 & 0 & 1.0 & 90.815 & 0.097  & 0.973\\ \hline
					90 & 99.344 & 0.007 & 1.0 & 100 & 0 & 1.0 &  90.347  & 0.101 & 0.972\\ \hline
					92 &  99.344  & 0.007 & 1.0 & 100 & 0 & 1.0 &  90.347  & 0.101 & 0.972\\ \hline
					\textbf{Average}&\textbf{99.453} & \textbf{0.007} & \textbf{1.0} & \textbf{98.828} & \textbf{0.012} & \textbf{0.998} & \textbf{87.191} & \textbf{0.122} & \textbf{0.953}\\ \hline
					\textbf{std-deviation}&\textbf{0.163}&\textbf{0.002}&\textbf{0}&\textbf{2.111}&\textbf{0.020}&\textbf{0.005}&\textbf{4.995}&\textbf{0.025}&\textbf{0.024}\\\hline
				\end{tabular}
				
			}
			\label{tab:malsyscall}
		\end{table*}
		
		\subsection{Evaluation of a set system calls employing large population test}
		\label{sec:lpt}
		As smartphones have limited computing resources, hence lightweight machine learning model is required to be installed on such devices. Keeping this in mind, we resort to applying two-step feature selection approach. Initially, system call set is synthesized by implementing Rough set-based feature selection. Subsequently, the attributes as mentioned earlier are further pruned by using statistical test, in particular, large population test, hence the method named as RSST.
		Two sample large population test are used to estimate if the population means differs~\cite{Massey2006}. Specifically, we apply a statistical test to determine set of system calls having increased divergence across the target classes. To carry this task, we consider same attribute set supplied to feature selection approach discussed in Section~\ref{sec:featureselection}. Specifically, the system call set used to build previous malicious model is further filtered using large population test. A similar approach is again repeated for feature set extracted from benign applications. The result of statistical method prunes around 50\% features. The significance is determined using a two-tailed test. Thus, the null hypothesis (H0) and the alternative hypothesis (H1) are defined as below: 
		\begin{itemize}
			\item \textit{Null Hypothesis (H0)}: The mean of system calls in malware and benign applications are the same.
			\item \textit{Alternate Hypothesis (H1)}: The mean of system calls in malware and benign set has significant difference.
		\end{itemize}
		
		\par The difference in mean of system call $i \in X$ is computed for both malware and benign set. The evidence of test is computed at the significance level $\alpha=0.05$ using equation~\eqref{eq:zscore}.
		
		\begin{equation}
			\label{eq:zscore}
			z=\frac{\overline{X}^i_M-\overline{X}^i_B}{\sqrt{\frac{\sigma^i_M}{|M|}+\frac{\sigma^i_B}{|B|}}},
		\end{equation}
		where $\overline{X}^i_M$ and $\overline{X}^i_B$, denote means of system call $i$ in the classes (malware/benign). Likewise, $\sigma^i_M$ and $\sigma^i_B$  are the standard derivations of system call $i$. The null hypothesis for a two-tailed test is rejected, if and only if, $z \leq 1.96$ and $z \geq 1.96$, indicating a significant difference in the mean of the system call in target classes.
		On examining the outcome of the result, 50\% of system calls having a small difference in means are excluded. In other words, the attribute space is constructed with the system calls that qualify the statistical test. Therefore, two feature list, one consisting of calls predominantly found in malware samples, and another set of dominant calls in the legitimate instances. Identical to previous experiments, evaluation metrics are measured with the variable amount of features, considered in increments of 10 system calls at a time. Overall 92 attributes were observed to satisfy $z-test$.
		
		\subsubsection{RSST-based Benign System calls}
		We observe from Table~\ref{tab:zscoreben} that Random forest provided an AUC value of 1.0 with a false positive rate of 0.001 at a feature length of 30. However, Rotation forest results in AUC value 1.0 with 30 system calls. AdaBoost again illustrates an FPR of 0.009 and AUC of 0.969 with 38 features~(found to comply statistical test). Average evaluation metrics show similar trends in the results for Random forest and Rotation forest. An important aspect to be noticed is the improvement in the performance of AdaBoost compared to previous experiments, as discussed in sections~\ref{sec:sigleg} and~\ref{sec:sigmal}, respectively.
		
		\begin{table*}[!htpb]
			\centering
			\caption{Result based on system calls invoked by benign applications obtained by applying RSST.}
			{
				\begin{tabular}{|c|c|c|c|c|c|c|c|c|c|}
					\hline
					\multirow{1}{*}{\textbf{Feature}} & \multicolumn{3}{c|}{\textbf{Random Forest}} & \multicolumn{3}{c|}{\textbf{Rotation Forest}}& \multicolumn{3}{c|}{\textbf{AdaBoostM1}}\\\cline{2-10}
					\textbf{Length}& \textbf{Acc} & \textbf{FPR} & \textbf{AUC} & \textbf{Acc} & \textbf{FPR} & \textbf{AUC} & \textbf{Acc} & \textbf{FPR} & \textbf{AUC}\\ \hline
					10 &  99.531 & 0.005 & 1.0 & 99.531 & 0.005 & 1.0 & 90.73 & 0.009 & 0.969\\ \hline
					20 &  99.625 & 0.004 & 1.0 & 99.906 & 0.001 & 1.0 & 89.7 & 0.107 & 0.967\\ \hline
					30 &  99.906 & 0.001 & 1.0 & 99.812 & 0.001 & 1.0 & 89.7 & 0.107 & 0.967\\ \hline
					38 &  99.812 & 0.002 & 1.0 & 99.906 & 0.001 & 1.0 & 90.637 & 0.009 & 0.969\\ \hline
					\textbf{Average}&\textbf{99.719} &\textbf{0.003} &\textbf{1.0}&\textbf{99.789} &\textbf{0.002} &\textbf{1.0}&\textbf{90.192} &\textbf{0.058} &\textbf{0.968}\\ \hline
					\textbf{std-deviation}&\textbf{0.171}&\textbf{0.002}&\textbf{0}&\textbf{0.177}&\textbf{0.002}&\textbf{0}&\textbf{0.569}&\textbf{0.057} & \textbf{0.001}\\\hline             
					
				\end{tabular}
			}
			\label{tab:zscoreben}
		\end{table*}
		
		\subsubsection{RSST-based Malware System calls}
		In Table~\ref{tab:zscoremal}, we observe that average evaluation metrics obtained with Random Forest and Rotation forest are approximately similar to the previous experiments. Random forest results in an AUC value of 1.0 with the false positive rate of 0.001 at a feature length of 30. While, Rotation forest gives an AUC value 1.0 with 20 system calls. AdaBoost demonstrates an average evaluation metrics compared with the results produced in sections~\ref{sec:sigleg} and~\ref{sec:sigmal}, interestingly with better a FPR.
		
		\begin{table*}[!htpb]
			\centering
			\caption{Performance based on system calls invoked by malware applications obtained by applying RSST.}
			
			{
				\begin{tabular}{|c|c|c|c|c|c|c|c|c|c|}
					\hline
					\multirow{1}{*}{\textbf{Feature}} & \multicolumn{3}{c|}{\textbf{Random Forest}} & \multicolumn{3}{c|}{\textbf{Rotation Forest}}& \multicolumn{3}{c|}{\textbf{AdaBoostM1}}\\\cline{2-10}
					\textbf{Length}& \textbf{Acc} & \textbf{FPR} & \textbf{AUC} & \textbf{Acc} & \textbf{FPR} & \textbf{AUC} & \textbf{Acc} & \textbf{FPR} & \textbf{AUC}\\ \hline
					10 & 99.221 & 0.019 & 0.998 & 99.532 & 0.005 & 1.0 & 81.361 & 0.176 & 0.874\\\hline
					20 & 99.532 & 0.005 & 1.0 & 99.906 & 0.001 & 1.0 & 87.828 & 0.121 & 0.937\\\hline
					30 &  99.906 & 0.001 & 1.0 & 99.813 & 0.002 & 1.0 & 89.513 & 0.107 & 0.955\\\hline
					38 & 99.625 & 0.003 & 1.0 & 99.906 & 0.001 & 1.0 & 90.637 & 0.009 & 0.969\\\hline
					\textbf{Average}&\textbf{99.571} & \textbf{0.007} & \textbf{1.0} & \textbf{99.789} & \textbf{0.002} & \textbf{1.0} & \textbf{87.335} & \textbf{0.103} & \textbf{0.934}\\ \hline
					\textbf{std-deviation}&\textbf{0.282}&\textbf{0.008} &\textbf{0.001}&\textbf{0.177}&\textbf{0.002}&\textbf{0}&\textbf{4.146}&\textbf{0.070}&\textbf{0.042}\\\hline              
				\end{tabular}
			}
			\label{tab:zscoremal}
		\end{table*}
		
		\par
\begin{figure*}[!htpb]
			\centering
			\includegraphics[scale=1]{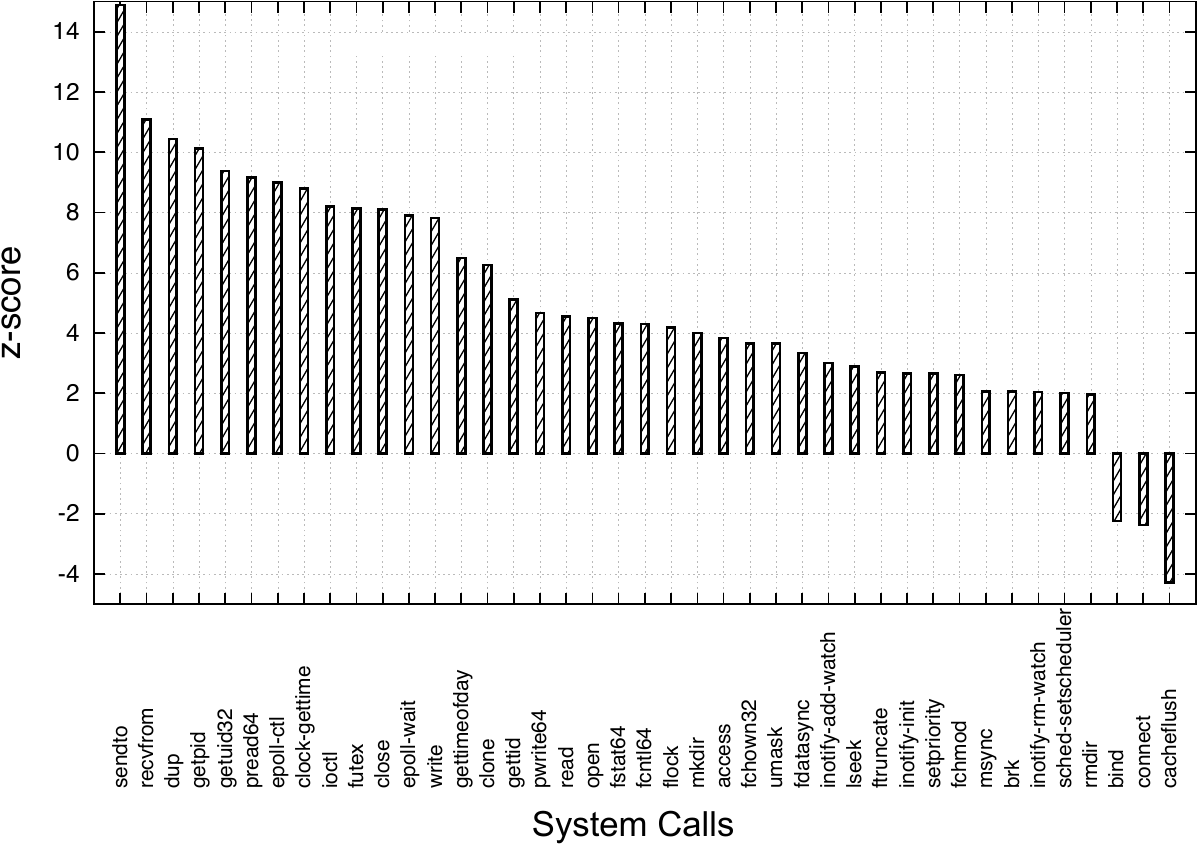}
			\caption{Discriminant System calls invoked by malware/benign applications.}
			\label{fig:prominentcalls}
		\end{figure*}

 Figure~\ref{fig:prominentcalls} shows the z-score value of prominent system calls participating in the system call space. These calls satisfy alternative hypothesis depicting substantial variance amongst feature vectors in target classes.
		
		\section{Comparative Analysis with Static and Dynamic Features}
		\label{comparative}
\begin{figure*}[!htpb]
			{\includegraphics[scale=0.30]{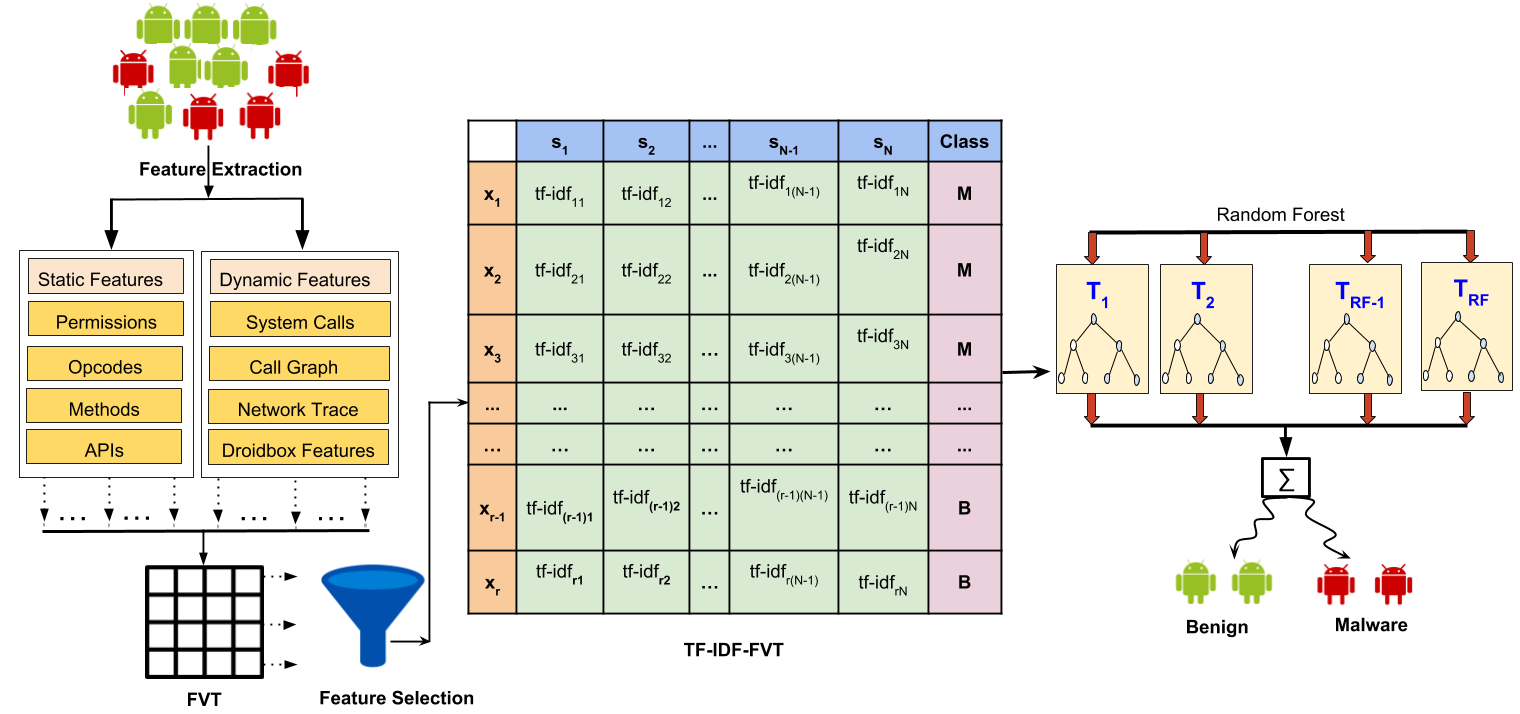}}
			\caption{Framework of malware scanner using static and dynamic features.}
			\label{fig:fullfeatures}
		\end{figure*}

		In order to validate the efficiency of system call feature set for identifying malicious instances, we further conducted series of experiments using diverse attributes/variables extracted using static and dynamic analysis. The framework of our experiment is shown in Fig.~\ref{fig:fullfeatures}. Here, we also estimate the result using F-measure along with the metrics used in all previous experiments. F-measure or F-score can be interpreted as weighted average of precision and recall, where low false positive rate indicates precision and low false negative rate relates to recall. F-measure reaches its best value at 1 and worst score at 0. The F-measure is harmonic mean of precision and recall. In most of the classification problem we have trade-off between precision and recall. If one of the parameters amongst precision and recall is favoured, the harmonic mean quickly decreases. However, F-measure is greatest when both precision and recall are equal. 
		
		\par Feature set using static analysis are formed by reversing \texttt{AndroidManifest.xml} and collection of \texttt{smali} files. In particular, attributes such as permissions, hardware components, app components, opcodes, and methods are considered. Additionally, features are collected by executing apks. Specifically, we derived attributes from \texttt{.pcap} files~(i.e., network-based features), system call graphs and information filtered using \textit{Droidbox}. The following section introduces aforementioned variables and the performance achieved by developing classification models incorporating them.
		
		\subsection{Evaluation on Static Features}
		\label{sec:comstaticfeatures}
		Static features are extracted from Android \texttt{Manifest.xml} and \texttt{smali} code of each applications. As discussed in previous experiments, each apk is transformed to a vector representation, which are used to create feature occurrence matrix. Later, prominent attributes are derived by applying feature selection approach as discussed earlier. The classification models are evaluated and the obtained results are shown in Table~\ref{tab:comparefeatures}.
		
		\begin{table}[h]
			\centering
			\caption{Performance with static features obtained at optimal feature length.}
			\begin{tabular}{|l|l|c|c|c|}
				\hline
				\textbf{Features} &	\textbf{Feature} &\textbf{TPR(\%)}	&\textbf{FPR} &\textbf{F1-measure}\\
				&\textbf{Length}& & &\\\hline
				Permission & 100 & 95.2 & 0.074 & 0.952\\\hline
				Activity &7000&93.4&0.142&0.932\\\hline
				Action &500 &90.6 &0.198 &0.903\\\hline
				Hardware&650 &79.9&0.529&0.74\\\hline
				Provider&800&72.5&0.70&0.617\\\hline
				Receiver& 5000&90.3&0.237&0.897\\\hline
				Service&3500&90.7&0.231&0.901\\\hline
				\textbf{Average}&-&\textbf{87.51}&\textbf{0.301}&\textbf{0.848}\\\hline\hline
				Malware(OPCODE) & 400 & 94.7 & 0.113 & 0.98\\\hline
				Benign (OPCODE) & 208 & 94.1 & 0.11 & 0.942\\\hline
				\textbf{Average}&-&\textbf{94.4}&\textbf{0.112}&\textbf{0.961}\\\hline\hline
				Malware(API) & 2550 & 91.2 & 0.189 & 0.908\\\hline
				Benign(API) & 1400 & 89.8 & 0.236 & 0.902\\\hline
				\textbf{Average}&-&\textbf{90.5}&\textbf{0.212}&\textbf{0.905}\\\hline\hline
				Malware(methods)&2500 &88.65 &0.061&0.919\\\hline
				Benign(methods)& 6500 &91.58& 0.145 & 0.909\\\hline
				\textbf{Average}&-&\textbf{90.12}&\textbf{0.103}&\textbf{0.914}\\\hline
			\end{tabular}
			\label{tab:comparefeatures}
		\end{table}
		
		\par Tabulated results demonstrate better performance for permissions comparing to the other statistical attributes. An application is not installed until a user accept all requested permissions. Developers may sometime declare permissions, which are not originally needed by an apk. Specifically, such applications are over-privileged and expose devices to threat. The top 5 permissions demanded by malicious applications are \texttt{INTERNET}, \texttt{READ\_PHONE\_STATE}, \texttt{WRITE\_EXTERNAL\_STORAGE}, \texttt{READ\_SMS} and \texttt{WRITE\_SMS}. 
		
		\par Machine learning system based on permissions can be defeated by having applications initially request fewer permissions during installation time. In particular, an adversary may create malicious applications to have an uniform statistical distribution of permissions as in benign dataset. Later, application(s) during execution may demand additional permissions. Under this scenario, the developed models will yield higher misclassification rate. Studies in~\cite{Moonsamy2012} report ex-filtration of sensitive data from the devices with the apps demanding zero permissions during installation. While authors in~\cite{Narain2016} illustrate zero permission app could be used to infer user's location, traveled routes using accelerometer, magnetometer and gyroscope. 
		
		\par External storage like SD Card contains sensitive data such as pictures, videos, configuration files, and backup documents, etc. Generally, applications have read-only access to the SD Card, allowing the attacker to fetch list of installed files. Alternatively, an adversary can query \texttt{/data/system/packages.list} to find list of installed applications, and subsequently determines exploits to compromise smartphones. Additionally, basic device information such as kernel version, device ID, and custom ROM can be distilled having access to \texttt{/proc/version} file.
		\par Table~\ref{tab:comparefeatures} shows the performance obtained by considering permissions. Extracted permissions are represented as binary vectors. The presence of a permission is denoted by 1 and absence by 0. With 100 significant permissions, an F1-measure of 0.952 with FPR of 0.074 is obtained. Permissions are ineffective for identifying malicious samples. In the outlook of end-user, list of permissions are generally viewed as a license agreement. It does not relate to the context of risk and neither indicates how much hazardous is the installed application. Besides, some permissions are frequently used by many applications thus the users do not care about them.
		
		\par Furthermore, analysis is performed by extracting instructions from each \texttt{smali} code. Generally, an instruction is composed of mnemonics and list of operands. In order to create feature set we considered mnemonics neglecting the operands. Prominent opcodes are selected with feature selection, two models are constructed: (a) one with relevant benign opcodes, and (b) another using malware opcodes. F1-measure of 0.98 was obtained with 400 malware opcodes (see Table~\ref{tab:comparefeatures}). We debate that opcodes/ngrams of opcodes cannot be effective in detecting unknown malware samples, as they can be easily obfuscated. Specifically, trivial obfuscation methods such as renaming of class/method/identifier can thwart detection. Consequently, scanners based on statistical signatures will imprecisely identify new samples.
		\par As an extension to the experiments, we extract API from malware and trusted applications. An F1-measure of 0.908 is obtained considering API's (see Table~\ref{tab:comparefeatures}). We thus argue that API's are weak attributes, as performance is inferior compared to permissions and opcodes. Present day malware employs reflection, so the applications refer malicious codes/libraries during execution. Malicious intentions are invisible during static analysis, as the set of API's in both malware and benign set appear identical. Since the distribution of attributes across feature vectors are likely to be uniform in the target classes, the classifier assigns incorrect labels to the apks.
		
		\subsection{Evaluation on Dynamic Features}
		\label{sec:comdynamicfeatures}
		To evaluate our scheme, we conducted experiments by executing applications. Further feature set are created using network trace, DroidBox information and system call graphs. The performance obtained with these features are compared with the set of system calls derived on applying the two-step feature selection approach (i.e., feature set created by applying rough set followed by large population test.). To practically show the efficiency of our scheme, the attributes and classification results are discussed in the following subsections.   
		
		\subsubsection{DroidBox attributes}
		DroidBox consists of two modules, one is the Host, and another is the Target. The Target inherits functionalities of TaintDroid, a dynamic taint analysis tool. It is launched from an emulator which monitors data at a low level. The Host part is a collection of Python scripts. The Host links emulator and receives information from the Target about the application being monitored. Finally, the outcome of the analysis is displayed in graphical or textual format. Few important information/sections retrieved from DroidBox are listed below:
		
		\begin{enumerate}
			\item \texttt{accessedFiles}-a list of files accessed by the application.
			
			\item \texttt{cryptousage}-operations associated with \texttt{cryptAPIAndroid.}
			
			\item \texttt{dataleaks}-gives user's data leak information.
			
			\item \texttt{fdaccess}-application performs read/write operations on files.
			
			\item \texttt{opennet/closenet}-open or close a socket.
			
			\item \texttt{recvnet/sendnet}-receive or transmit via network.
			
			\item \texttt{sendsms/phonecall}-send sms or call specific number.
		\end{enumerate}
		\par The DroidBox tracing file is a record of actions in JSON format. Largely all sections in the JSON file have the following format.
		
		\begin{verbatim}
		"Section name":{
		"Time of operation"{
		"Parameter (e.g., for accessedFiles,î 
		ìpath of the accessed file)
		}
		}
		\end{verbatim}
		\par Consider an example shown below, the application being monitored repeatedly accessed same files, i.e., \texttt{abc.png} and \texttt{imagesñ12345ñ54321-example.jpg}, may indicate suspicious activity.
		
		\begin{verbatim}[formatcom=\sffamily]
		ìaccessedFilesî: {
		ì100010001: ì/mnt/sdcard/Download/abc.jpgî,
		ì1000112345î: ì/mnt/sdcard/Download/î
		ìimagesñ12345ñ54321-example.jpgî,
		ì1000112450î: ì/mnt/sdcard/Download/abc.jpgî,
		ì1000113500î:  ì/mnt/sdcard/Download/î
		ìimagesñ12345ñ54321-example.jpgî,
		&lt;Ö&gt;
		}
		\end{verbatim}

		\par Using the section and its associated fields/parameters, we represent each app vector in the form of integers~(i.e., occurrence of operation along with its parameters). Table~\ref{tab:dynamicfeatures} depicts the performance obtained with DroidBox.
		
		\subsubsection{Network trace}
		Network traffic are extracted using \texttt{tcpdump}~\cite{tcpdump} after installing application in emulator. Like the earlier experimental setting used for system call analysis, Android Monkey is used to interact with the application. The network traffic is recorded until fixed random event existed. Finally, the output of network trace is recorded in a \texttt{.pcap} file. The basic structure of \texttt{tcpdump} output is shown below:\\
		\{timestamp\} \{network protocol\} \{source ip\}.\{source port\} $>$ \{dest ip\}. \{dest port\}
		
		\par Subsequently, we extract six features and classification model is built. An AUC of 0.994 with an FPR value of 0.033~(refer Table~\ref{tab:dynamicfeatures}) is obtained using six features listed below:
		\begin{enumerate}
			\item \textit{Raw traffic size~(RS)}: is the total packet size estimated at the time frame of analysis.
			\item  \textit{Number of packets(PN)}: count of packets in a pcap file. 
			\item  \textit{Average length of packet(AL)}: average length of packets for each pcap file.
			\item  \textit{Outbound packets in a file(ON)}: is number of packets transmitted from a source IP to the others.
			\item  \textit{Inbound packets in a file(IN)}: number of packets received by a source IP.
			\item  \textit{Total outbound and inbound packets(OIN)}: sum total of inbound and out-band packets.
		\end{enumerate}
		
		\subsubsection{System call graph}
		\label{sec:comsystemcallgraph}
		In order to ascertain relationships among the logged system calls, a directed graph $G=<V,E>$ is constructed, where $V$ is the set of vertices and $E$ is the set of edges. In particular, the graph is represented in the form of the adjacency matrix. For each pair of extracted system calls, edge between the vertices are created. The weight of edge is incremented for each appearance of system call pairs. Subsequently, we compute in-degree, out-degree, standard deviations of in-degree/out-degree, which is used as features for building the classification model. Table~\ref{tab:dynamicfeatures} exhibits the classifier outcome with graph based features. We noticed that detection performance of graph-based features are better than attributes obtained from network trace and DroidBox. This indicates that the interactions of an application with the operating system~(using system calls) definitely appears to be strong candidate for developing malware detection system.       
		\begin{table}[h]
			\centering
			\caption{Results with features obtain with dynamic analysis.}
			\begin{tabular}{|l|l|c|c|c|}
				\hline
				\textbf{Features} &	\textbf{Feature} &\textbf{TPR(\%)}	&\textbf{FPR} &\textbf{AUC}\\
				&\textbf{Length}& & &\\\hline
				DroidBox&25 & 91.7&0.083&0.983\\\hline
				Network&6&96.1&0.033&0.994\\\hline
				System call graph&350& 97.45&0.224& 0.994\\\hline
			\end{tabular}
			\label{tab:dynamicfeatures}
		\end{table}     
		
		\section{Performance with Conventional Feature Selection Approach}
		\label{sec:performanceconventional}
		In this section, we briefly introduce well-known feature selection algorithms used in the domain of malware analysis. These algorithms select a subset of system calls which typically output enhanced classification rate. In particular, feature selection algorithms: Information Gain(IG), Chi-Square (CHI), Correlation-based Filter (CFS) and Wrapper Subset Evaluator (WSE) are utilized to select relevant attributes before developing classification model. We have selected open source implementation of presented algorithms included in WEKA. Eventually, the detection performance of our proposed two-step feature selection model is compared with the models developed from aforementioned feature selection algorithms.
		
		\par \textit{Information Gain~(IG)} is based on the concept of information theory. In this method, the algorithm calculates the amount of information carried by as system call $(s)$. $IG(s)$ involves computing entropy of a class $H(C)$ and subtracting the conditional entropy of $s$ after observing the class, i.e., $H(s|C)$. Hence, for a classification system, $IG(s)$ is expressed using equations~\eqref{eg:ig}-\eqref{eg:ig2}.
		
		\begin{equation} \label{eg:ig}
			IG(s) = H(C)-H(s|C), 
		\end{equation}
		\begin{equation} \label{eg:ig1}
			H(s)  = -\sum_{s\in\,A}p(s)\log_2(p(s)),
		\end{equation}
		\begin{equation} \label{eg:ig2}
			H(s|C)  = -\sum_{\{M,B\}\in\,C}p(C)\sum_{s\in\,A}p(C|s)\,\log_2(p(C|s)),
		\end{equation}
		Finally, $IG(s)$ of all system calls are arranged in descending order, and the top system calls are used for modeling. \textit{Chi-Square(CHI)} feature selection is used to test the independence of two events. In our case, the two events are occurrence of a system call and presence of the class. Precisely, we want to evaluate whether the occurrence of a system call and class are independent. Our aim is to determine set of calls such that its presence and class are highly dependent. Importance of a system call $s$ is calculated using equation~\eqref{eq:chisquare}.
		\begin{equation}
			\chi^{2}\,(s) = \frac{N(AD-CB)^2}{(A+C)\times(B+D)\times(A+B)\times(C+D)},
			\label{eq:chisquare}
		\end{equation}
		where $N$ is the total number of apks ($N=A+B+C+D$), $A$ and $B$ are the number of malware and benign applications containing the system call $s$, $C$ and $D$ are number of malware and benign instances without $s$. Like $IG(s)$, system calls are sorted based on $\chi^2$ value, finally we select the top ranked system call for developing classification model. \textit{Symmetric Uncertainty~(SU)} neutralizes the bias induced by IG towards higher values and normalizes it in range of [0,1]. Symmetric Uncertainty is the measure of information contained in variables $A$ and $C$ put together over the information independently contained in $A$ and $C$. The value 1 indicates that knowledge of one variable can determine another attribute. Additionally, it denotes two variables are highly correlated. On the other hand value 0 signifies independence of variables. Symmetric Uncertainty is defined as:
		\begin{equation}
			SU(A,C) =2.\frac{IG(A|C)}{H(A)+H(C)}.
		\end{equation}
		System calls having higher correspondence with the class are used for developing classification models. \textit{CfsSubsetEval} is a correlation-based feature selection approach. The algorithm selects predominant attributes/system calls based on two aspects (i) correlation of an attribute and class must be high. It assures the relevance of system call and a class($M/B$), and (ii) the set of system calls obtained from the previous step must not have high correlation amongst each other (higher correlation means larger redundancy). In other words, features/calls are effective if its correlation with class is large, and all its redundant groups are discarded. \textit{Wrapper Subset Evaluator~(WSE)} looks for attributes along-with the given classifier. Hence, in the process of finding subset of calls, certain search mechanisms are used. Hence, we employed two well known search approach i.e., Genetic search~(GS) and Breadth First Search~(BFS) respectively.
		
		\par
\begin{figure*}[!htpb]
			\centering
			\begin{subfigure}{0.325\textwidth}
				\centering
				\includegraphics[width=1\textwidth]{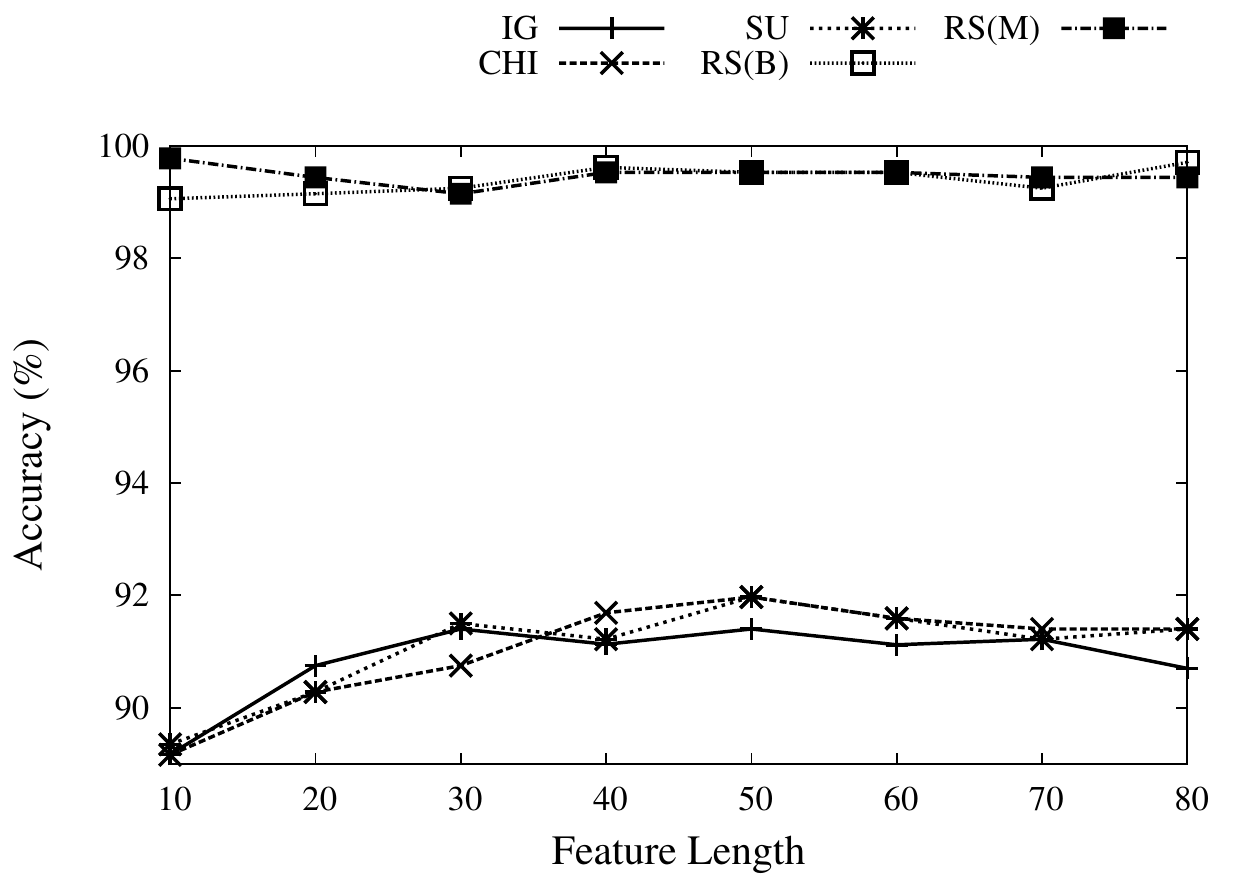}
				\caption{ACC vs. feature length}
				\label{fig:ACCplot}
			\end{subfigure} 
			\begin{subfigure}{0.325\textwidth}
				\centering
				\includegraphics[width=1\textwidth]{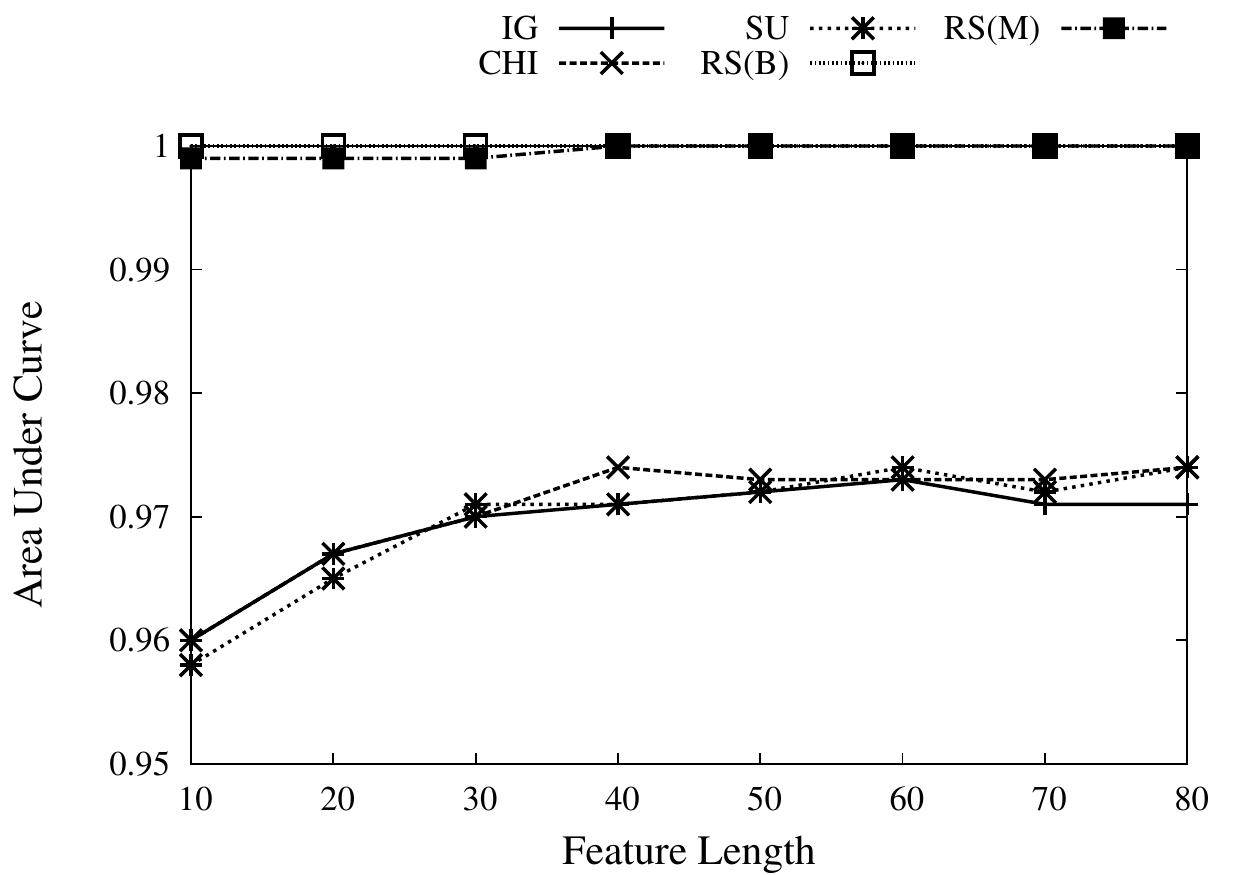}
				\caption{AUC vs. feature length}
				\label{fig:AUCplot}
			\end{subfigure} 
			\begin{subfigure}{0.325\textwidth}
				\centering
				\includegraphics[width=1\textwidth]{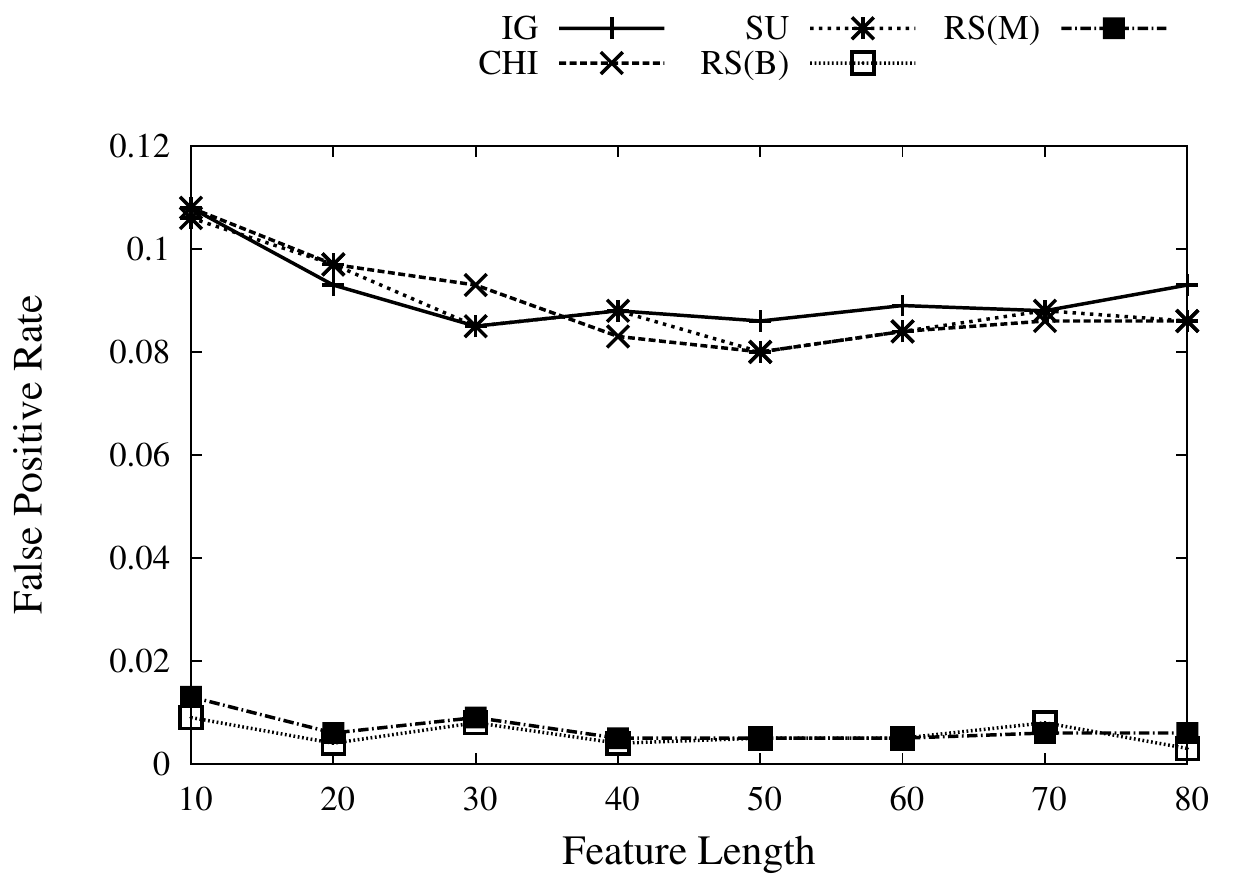}
				\caption{FPR vs. feature length}
				\label{fig:FPRplot}
			\end{subfigure} 
			\caption{Comparisons of feature selection methods on ACC, AUC and FPR over various length of selected features. IG: Information Gain; CHI: Chi-Square; SU: Symmetrical Uncertainty; RS~(B): Rough set-benign system calls; and RS~(M): Rough set-malware system calls.}
			\label{fig:fig4}
		\end{figure*}

 After applying the aforementioned feature selection methods, list of relevant system calls are obtained. These set of system calls constitute our feature set. The performance of classification is determined by varying the length of features. We can see clearly that high accuracy and AUC is obtained with our proposed feature selector on comparing IG, CHI, SU, CFS, and WSE~(both BFS and GS search techniques). Figures~\ref{fig:ACCplot}-\ref{fig:FPRplot} show the achieved outcomes. In particular, accuracies obtained with conventional approaches are between 89-92\%, AUC is in range of 0.96-0.97 and FPR lies between 0.08-0.11 respectively. Since the implementations of algorithms: CFS and WSE in WEKA returns a single subset of system calls, we estimated the performance of the models on these subsets. CFS reported an accuracy of 90.28\%, with FPR of 0.0971 and AUC of 0.966 at 14 attributes. WSE~(BFS search) resulted in 79.16\% accuracy with 0.241 FPR and 0.832 AUC at with two significant calls. Additionally, using WSE~(GS search) we obtained 84.4\% accuracy with an FPR of 0.157 and AUC value of 0.918.
		\par We note that, the performance of classification models created from conventional feature selection approaches are not better. Hence, performance assessment of combined system call set is undertaken. The combined feature set include collection of attributes from individual classification models which is supposed to produce improved results. In particular set of 16, 17, 30, 50 and 50 significant calls filtered with WFS (GS), CFS, IG, CHI, and SU are grouped together. Finally, we obtain 54 unique calls by combining the previous outputs. Specifically, the aforementioned attribute set is created to retain system calls with reasonable predictive capabilities from several models in order to achieve increased detection. Eventually, the best outcomes of this experiment are 91.5\% accuracy, with 0.085 FRP and 0.972 as AUC. We thus conclude that combined feature space does not improve detection as opposed to selectors independently considered. To validate this, the feature set is inspected and we noticed it to be augmented with irrelevant calls. Hence a comprehensive approach for feature fusion~\cite{ruta2000overview} is required to be investigated, which is not within the scope of the present study, and will be considered in the future experiments.
		
		\section{Discussion}\label{discussions}
		In this section, we discuss important conclusions drawn based on the investigations conducted through multiple experiments.  Particularly, the inferences are summarized on the following interpretations:
		\begin{enumerate}[(1)]
			\item The performance of detection reported with the feature set comprising of system calls, call graph attributes and network traces are competitively the best compared to the static features. An app may utilize reflection and native code~\cite{felt2011android, rastogi2013droidchameleon} to make its real program logic undetectable by static analysis. One of the important conclusions drawn from extensive experiments is that behaviour of malicious and benign samples are appropriately represented with attributes derived from the dynamic analysis. Static analysis is not resilient to typical obfuscation techniques~\cite{rastogi2014catch1}. Statistics from~\cite{dong2018understanding} report interesting facts like 43\% of Google Play Store apps are obfuscated, 73\% of third-party markets and 63.5\% of malicious apps use identifier renaming to obfuscate applications. Moreover, the same study demonstrated that malware authors employ string encryption to hide true intentions of the malicious code, which is rarely observed in legitimate applications. Besides, source code can be conveniently altered using ProGaurd~\cite{progaurd}, an obfuscation tool. Additionally, solution based on static control flow analysis can be defeated by adopting DashO~\cite{dasho} a Java and Android obfuscator. Thus, the machine learning based solution depending on static features is supposed to give higher misclassification rate. Studies in~\cite{dong2018understanding} state that trusted applications are equally obfuscated as with malicious counterparts. This is performed in order to optimize the bytecode and protect benign apks against code reversing attacks. The aforesaid, obfuscation techniques do not affect the performance of machine learning approaches utilizing system calls and features derived from system call graphs.      
			
			\item It is evident from Fig.~\ref{fig:fig5} that the average values of evaluation metrics obtained with proposed feature selection method using statistical test following rough set are better compared with commonly employed approaches. The proposed attribute selector initially determines a subset of system calls with high-class dependency/significance. Subsequently, the large population test filters irrelevant system calls to obtain a subset of system calls having a significant mean difference across the classes to generate a reduced system call set.    
			\begin{figure*}[!htpb]
				\centering
				\begin{subfigure}{0.325\textwidth}
					\centering
					\includegraphics[width=1\textwidth]{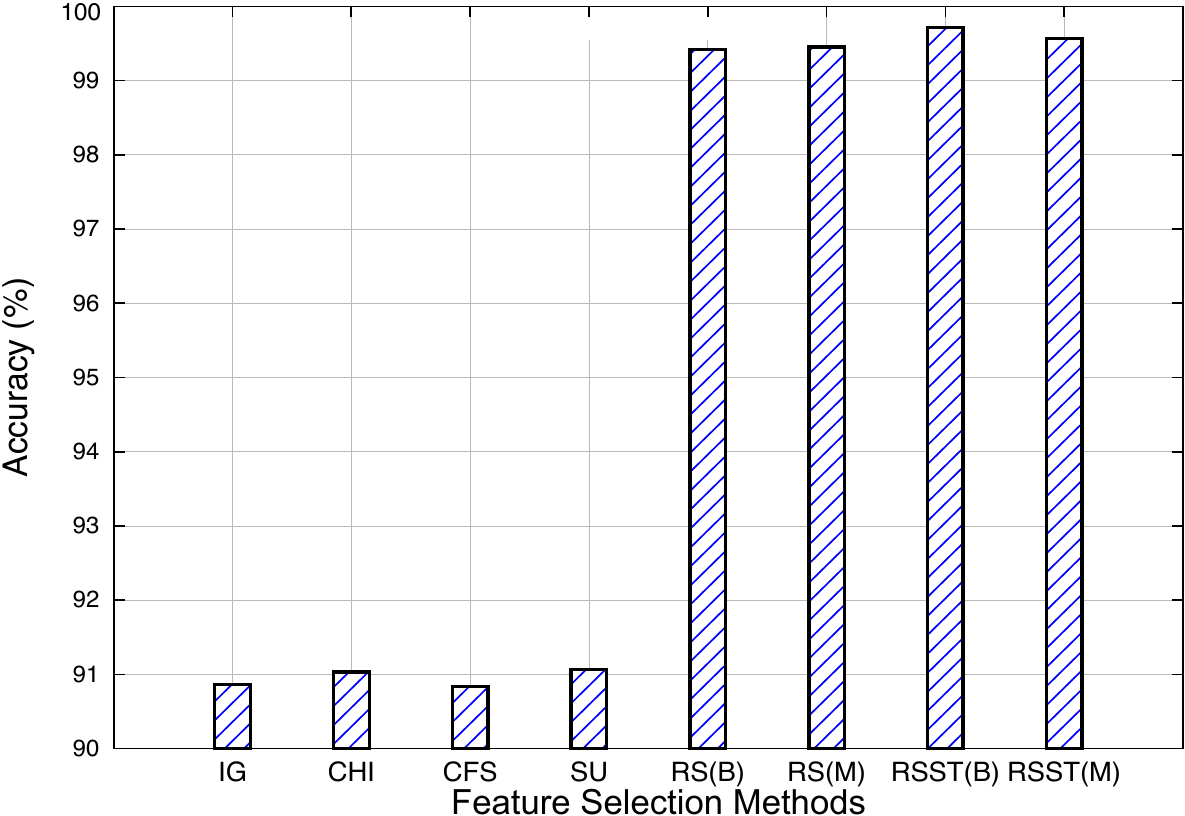}
					\caption{Average ACC}
					\label{fig:AVGACCplot}
				\end{subfigure} 
				\begin{subfigure}{0.325\textwidth}
					\centering
					\includegraphics[width=1\textwidth]{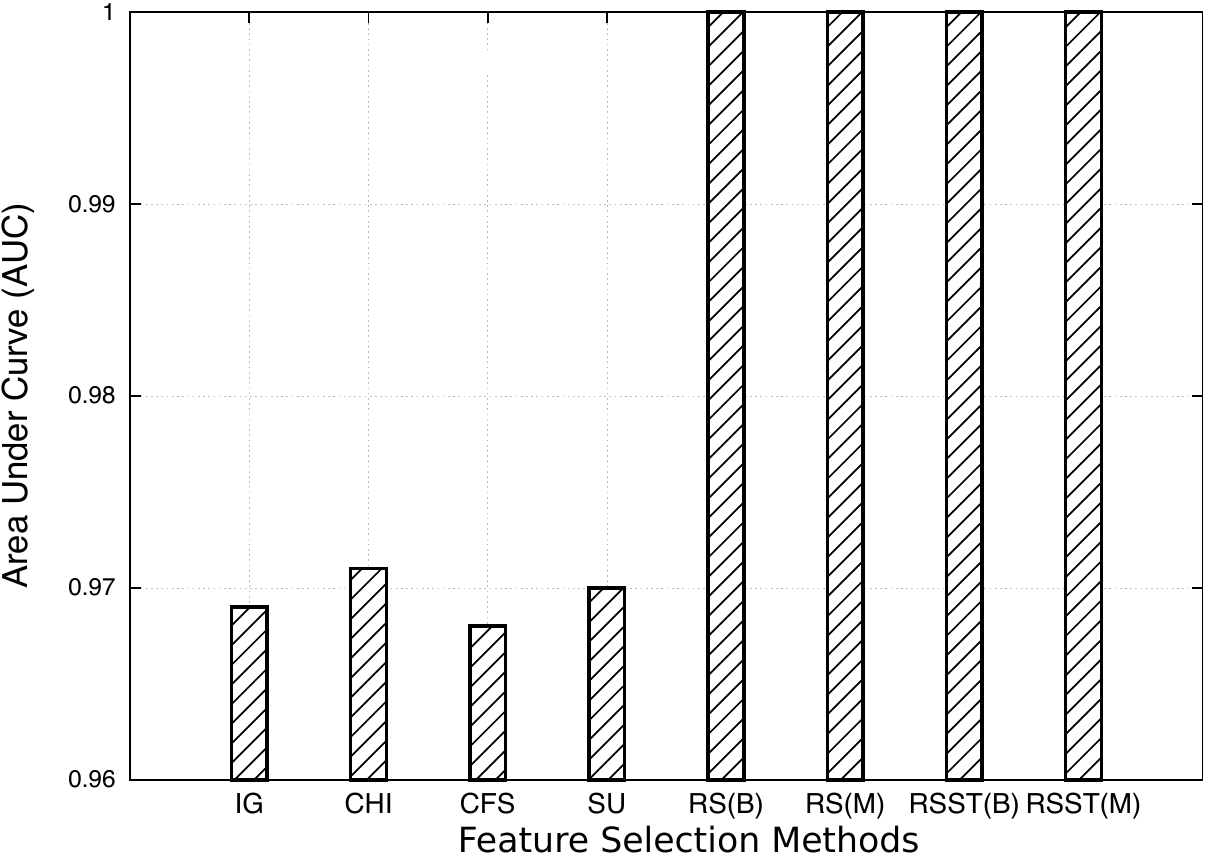}
					\caption{Average AUC}
					\label{fig:AVGAUCplot}
				\end{subfigure} 
				\begin{subfigure}{0.325\textwidth}
					\centering
					\includegraphics[width=1\textwidth]{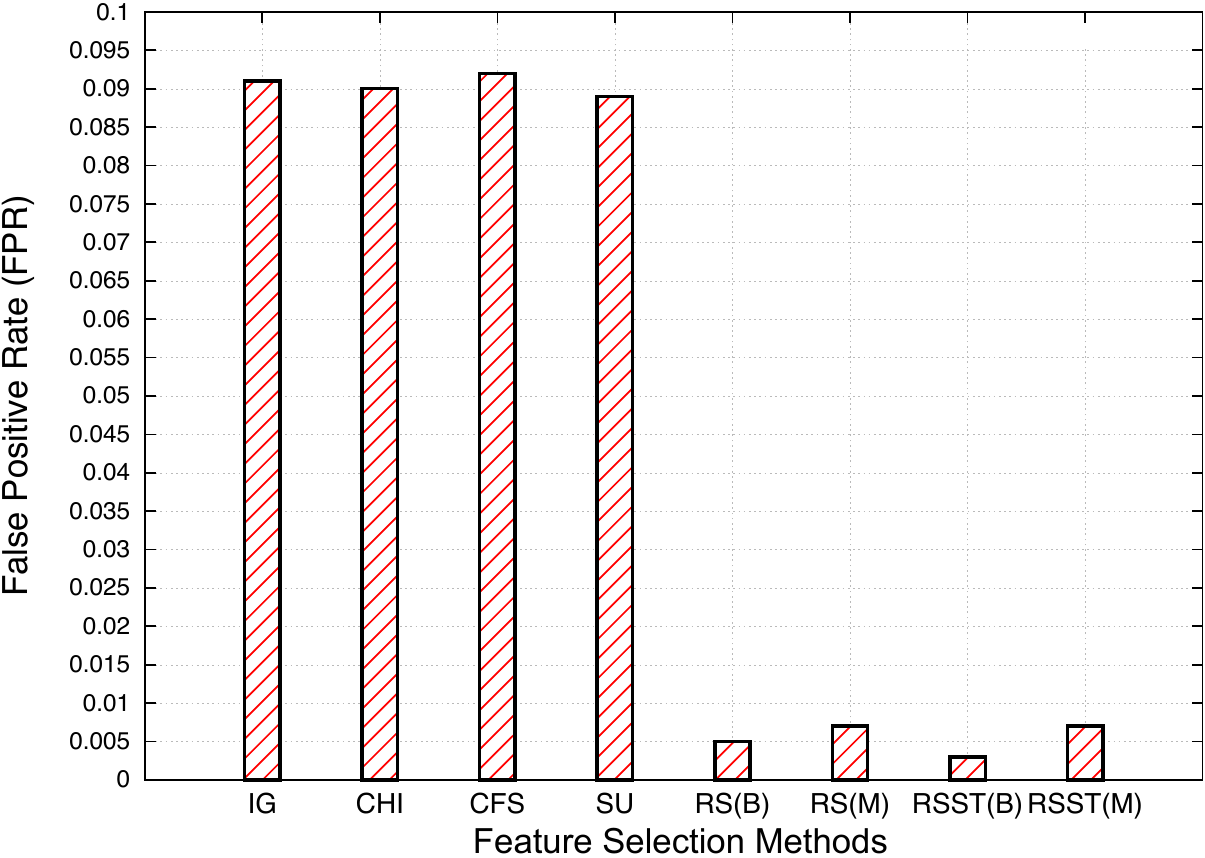}
					\caption{Average FPR}
					\label{fig:AVGFPRplot}
				\end{subfigure} 
				\caption{Comparisons of average performances of various feature selection methods. IG: Information Gain; CHI: Chi-Square; SU= Symmetrical Uncertainty; RS~(B): Rough set-benign system calls; and RS~(M): Rough set- malware system calls.; RSST~(B): benign attributes obtained after two-step feature selection i.e., Rough set followed by large population test; and RSST~(M): malware features collected by adopting two-step feature selection, i.e., Rough set followed by large population test.}
				\label{fig:fig5}
			\end{figure*}     
			
			\item Exhaustive experiments performed by us demonstrate that the feature set comprising of few system calls lack its representativeness about the target classes. In other words, fewer attributes are incapable of exhibiting separation~(or numerical variance) between feature vectors. Thus, the detection rate is also less. This conclusion can be clearly drawn from Fig.~\ref{fig:fig4} especially until 30 system calls. Consequently, the accuracy and AUC are small with large FPR. Furthermore, the addition of calls in the attribute set increases the difference in feature vectors. Hence, the prediction capability of classifier also improves, this is visualized for the feature set comprising of 30-50 system calls. However, augmenting the feature set with additional calls result in the classifier to learn attributes that add noise to the feature set. Hence, the performance remains steady and gradually tend to decrease. This trend can be perceived for feature-length beyond 50 attributes. The decrease in the performance is primarily due to the increased variability across the samples resulting in the wrong prediction.
			
			\item It has been observed that like any security systems/solutions, machine learning based malware detectors can be bypassed by adversarial examples. In general, the adversarial samples are crafted by adding small perturbation so that these examples are misclassified by learned models. In particular, the adversarial samples are crafted to match the distribution of attributes in the malicious and legitimate set. Normally the classification models developed with permissions and APIs are learned with applications represented in binary values, with the presence of feature indicated by one and absence is shown with zeros. The perturbations are induced by changing values of some permissions/APIs with zero elements to ones. These extraneous attributes are included to evade anti-malware systems. Specially, the permission-based detectors can be easily bypassed by augmenting \texttt{AndroidManifest.XML} files with extra permissions. 
			\par Machine learning models created with system calls are difficult to be bypassed with the adversarial examples. To do this, the malware developers must learn the statistical difference of each system call from a large collection of surrogate data set. Later, the source code of each app must be modified to include certain functionality resulting in the invocation of calls that match the distribution of calls in the benign set. However, it is difficult to be achieved, as it would require malicious apps consume more execution time. Besides, such samples can be identified \textit{without} substantial effort employing trivial heuristics such as (a) battery used; (b) percentage of CPU utilized; (c) amount of free memory available; and, (d) calls to extraneous system calls~(e.g., get current date/time, list of files/threads/processes, etc.).
			
			\item We estimated the time required in extracting static features from \texttt{AndroidManifest.xml} and smali code. The total time spent to extract manifest features~(permissions, services, activities, hardware etc.) is observed as 1073 seconds. Hence, per sample time on an average is estimated to be 0.3256 seconds. Also, time invested to gather API and opcodes is determined as 3229 seconds. Thus on an average, time spent for a single instance is 0.922 seconds. Since dynamic analysis involves execution of an application in an emulated environment, the average time required to extract system calls is estimated as approximately 3 minutes. Time invested to construct classification model and later generating results using cross-validation is between 0.425-0.577 seconds for all feature selectors considered in this paper. The minimum time is obtained for the model developed with CFsSubsetEval, and maximum time of 0.577 seconds is invested for models created using RSST, learned with malware system calls.
			
			\item Looking at the performance of classification algorithms, in context of identifying samples and running time, we conclude, the effectiveness of both Random forest and Rotation in detecting malware samples with the high classification accuracy, i.e., 99.9\%. However, on average, experiments with AdaBoost resulted in the best accuracy of 90.19\%  which is far less than other two algorithms. Essentially Random forest and Rotation forest assigns the class label of a sample using majority voting. On examining the evaluation metric, we conclude that a fair selection amongst Random and Rotation forest cannot be made. However, we preferred Random forest over Rotation forest, as the time required for building classification model with Random forest was observed to be less compared to Rotation forest, refer to Figure~\ref{fid:runningtimes}. Furthermore, the running times for Rotation forest increases with the size of attributes set, similar trends were also observed by author in~\cite{du2015random}.
			
			\begin{figure}[h]
				\centering
				\includegraphics[scale=0.75]{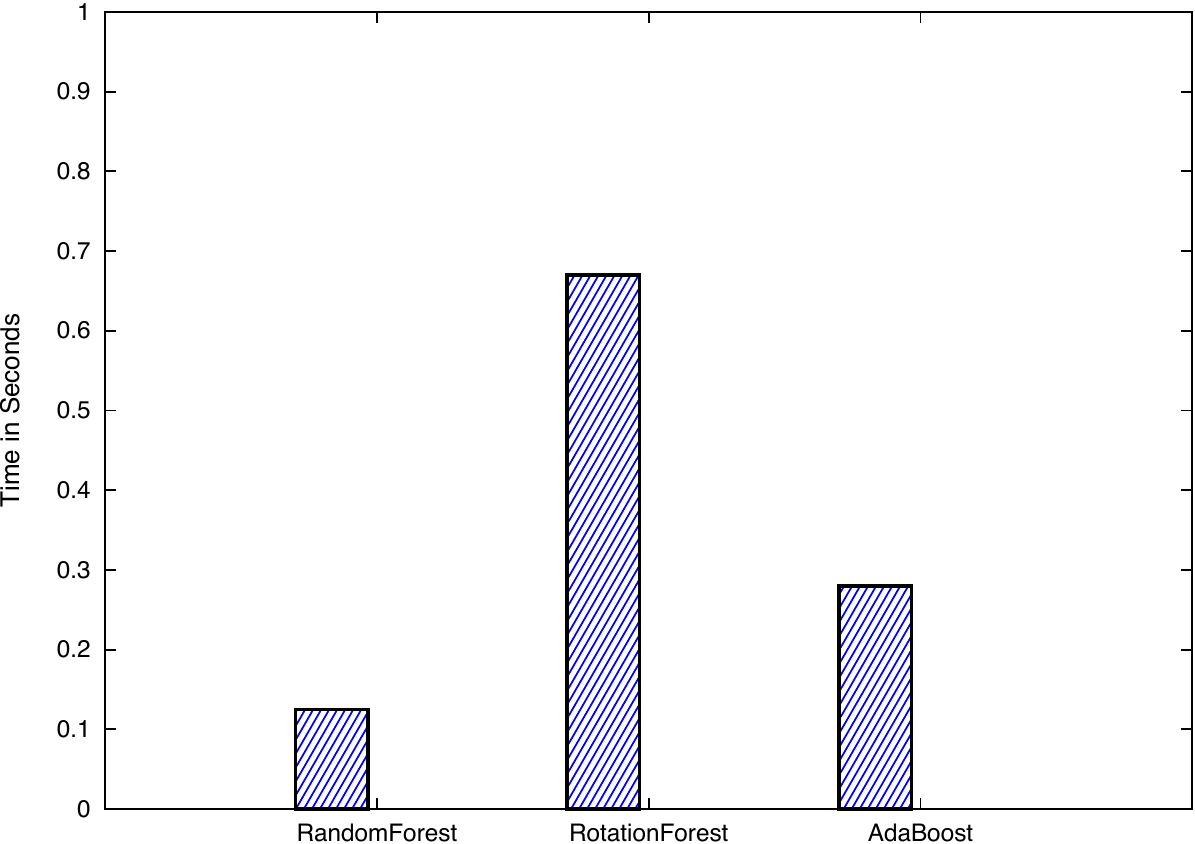}
				\caption{Average running times of classification algorithms.}
				\label{fid:runningtimes}
			\end{figure}
			
		\end{enumerate}
		
		\section{Comparison with Prior Works}
		\label{sec:comparepreviouswork}
		In this section, we compare the results of our proposed work with earlier studies. Authors in~\cite{damshenas2015m0droid} proposed an Android malware detector known as M0Droid consisting of two modules client agent and server analyzer. The agent executes in the background and submits the application to the server, which executes an application in the emulator. During execution, each apk is subjected to random events generated by Android Monkey, consequently, system calls are recorded. Later, a signature for each application is created, which is represented as a sequence consisting of a pair of system call identifier and occurrence. Detection of unknown instances is undertaken by comparing the Spearman correlation coefficient with known signatures in the repository. The experimental study demonstrated 60.16\% detection rate with 39.43\% FPR on 200 applications.
		
		\par Xi Xiao~\cite{xiao2016back} \textit{et al.} considered 196 system calls. To achieve improved classification accuracy, back propagation neural network on Markov chains from system call sequence were considered. Each system call was treated as a state thus, corresponding to 196 states. Android Monkey was used to generate 1000 pseudo-random events and later, \texttt{strace} was used to record system call events. The analysis was performed to determine optimal network structure (i.e., depth, hidden--layer and learning parameter), deployed for identifying unseen malware instances. Thus, assessment of three and four layers neural network was carried away. Three hidden layer with 37 nodes resulted in highest $F-score$ of 0.980 at a $TPR$ of 0.974 and $FPR$ of 0.0134. However, with four-layer network better performance was achieved when the first hidden layer was between 450--650. The highest $F-score$ of 0.982 at $TPR$ of 0.977 and a $FPR$ value of 0.013 was reported. Moreover, better classification outcome was ascertained with 0.7 as learning rate. The methodology suffers from certain limitations (a) on varying kernel version, the number of system call entries increases. Thus, classification model with fewer discriminative calls must be determined, otherwise, noisy attributes might result in over parameterization leading to poor generalization and (b) more the number of system calls, training will become excessively time consuming. 
		
		\par In~\cite{canfora2015detecting} system call sequences were considered, and insignificant calls were eliminated by determining \textit{relative class difference} for a system call. Execution trace of an application was represented using boolean values. The study also considered estimation of mutual information of a system calls with respect to class labels to ascertain calls representative of a target class. Subsequently, a feature to feature correlation was also estimated which was reported to deliver poor results. A malware detection accuracy of 97\% was obtained with this approach. The principal limitation we observed with feature selection approach implemented in paper is the absence of association of class weight with the occurrence of system call.
		\par In~\cite{zhao2015fest}, authors proposed Feature Extraction and Selection Tool~(FEST) for malware detection. The feature extraction module was designed to extract permissions and APIs. Subsequently, prominent attributes were picked using the proposed feature selection algorithm, \textit{FrequenSel}. The study reported an accuracy of 98\%, with 2\% false positive rate. We implemented a similar version of feature selection algorithm~(i.e., FrequenSel) on the set of system calls extracted from our dataset, and obtained an accuracy of 90.43\%, with 8.9\% FPR. Also, with 25 and 6 significant system calls independently extracted from benign and malware dataset, 89.916\%, and 89.635\% accuracy is obtained, with 9\% and 8.9\% FPR.
		\begin{table*}[!ht]
			\centering
			\caption{Comparative Analysis with Prior Works.\vspace{-10px}}
			\begin{tabular}{|p{1.2cm}|p{4.5cm}|p{3cm}|p{4.5cm}|}\hline
				\textbf{Prior} & \textbf{Approach} & \textbf{Metrics}&\textbf{Remarks}\\
				\textbf{Works} & & & \\\hline
				\cite{damshenas2015m0droid}&Dynamic analysis, system call identifier along with frequency &Detection rate 60.16\%, 39.43\% FPR& Analysis performed on limited set of apks. Attribute selection method is not employed. Increase in threshold produced high FNR\\\hline
				\cite{xiao2016back} & 196 system calls & $F-score$ is 0.982, $TPR$ is 0.977 \& $FPR$ is 0.013 & Malicious call sequence can be disturbed by injecting system calls found in trusted samples. Training is expensive. \\\hline
				\cite{canfora2015detecting} & System call sequences &accuracy is 97\% & Feature selection method does not employ weight of class along with score assigned to system call. \\\hline
				\cite{zhao2015fest} & Permissions and APIs & 98\% accuracy, with 2\% FPR & Same algorithm on system calls resulted in 90.43\% accuracy with 8.9\% FPR. The chosen threshold cannot determine calls that clearly separates the target classes.\\\hline
				\textbf{Our Method} & System calls & 99.9\% accuracy with 1\% FPR using 30 calls & the proposed feature selection algorithm outperforms conventional selectors discussed in Section~\ref{sec:performanceconventional}. \\\hline
			\end{tabular}
		\end{table*}
		
		\section{Limitations and Future Direction}
		\label{limitandfuturework}
		Our approach carries the general weaknesses of supervised learning model. The proposed method utilizes training data to build the model, under the assumption that the testing data, which the model will be applied to, is drawn from the same population as the training data. This assumption is not true in reality, since the malicious application may evolve. Hence, the model needs to be updated with new training data including new benign and and malicious applications. The extension of our work  will consider the system call sequences and call graph to construct semantic features. These attributes will be used to classify malware instances to different families. To this end, we may consider creating an iterative classification system. In this scheme, we plan to develop multi-tier classification approach. Instances will be initially classified using a set of classifiers and feature selection methods. Misclassified samples from previous layers will be further supplied in the subsequent layers employing diverse classifier and feature selectors. The features from each layer will be subsequently combined to generate final attribute set which will be used for modeling and prediction. 
		We would like to extend the aforementioned approach employing deep learning methods.
		\par Finally, we intend to extend our study by evaluating the approach on multiple dataset. Besides, we plan to investigate the robustness of classification system by incorporating attribute fusion approach by coupling call sequences with other features derived from source code of applications. 
		
		\section{Conclusions}
		\label{conclusion}
		In this paper, we propose a two-step feature selection approach utilizing predictive capabilities of rough set and a statistical test for determining relevant system calls. We observed that with 30 significant system calls we could separate malware and goodware~(i.e., non-malicious) with 99.9\% accuracy, AUC of 1.0, and 1\% FPR. Comprehensive analysis of the proposed feature selector with conventional feature selection methods such as Information Gain, Symmetric Uncertainty, ChiSquare, and CFsSubsetEval has been performed to test the performance. The results demonstrate that the proposed feature selection algorithm outperformed the traditional techniques. Exhaustive experiments with static attributes derived from manifest files, smali code, and features obtained using dynamic analysis including network traces, call graph attributes and droidbox information, suggest that feature set comprising of system calls exhibited better performance.

		\section*{Acknowledgments} \label{sec:9}
		This work is also partially supported by the grant n. 2017-166478 (3696) from Cisco University Research Program Fund and Silicon Valley Community Foundation, and by the grant "Scalable IoT Management and Key security aspects in 5G systems" from Intel. Moreover, the work is supported by the project ``Adaptive Failure and QoS-aware Controller over Cloud Data Center to Preserve Robustness and Integrity of the Incoming Traffic'' funded by the University of Padua, Italy.

		\bibliography{bibliography}
		\newpage
		
		\section*{Author Biography}
		
		\begin{biography}{\includegraphics[width=75pt,height=75pt]{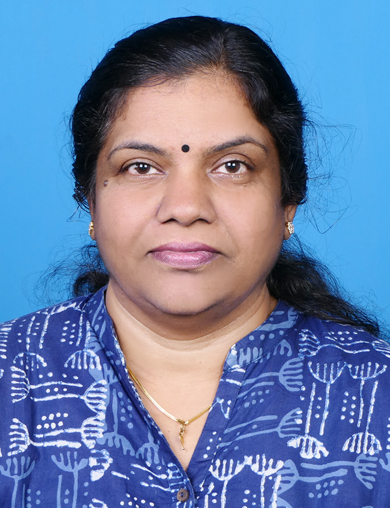}}{\textbf{Deepa K.} is currently persuing her Ph.D from Bharathiar University, Coimbatore. She is MCA \& MPhil in Computer Science from Bharathiar University. Her main research interests Cyber Securities and Machine Learning. Deepa K. is currently persuing her Ph.D from Bharathiar University, Coimbatore. She is MCA \& MPhil in Computer Science from Bharathiar University. Her main research interests Cyber Securities and Machine Learning.}
		\end{biography}
		
		\begin{biography}{\includegraphics[width=75pt,height=86pt]{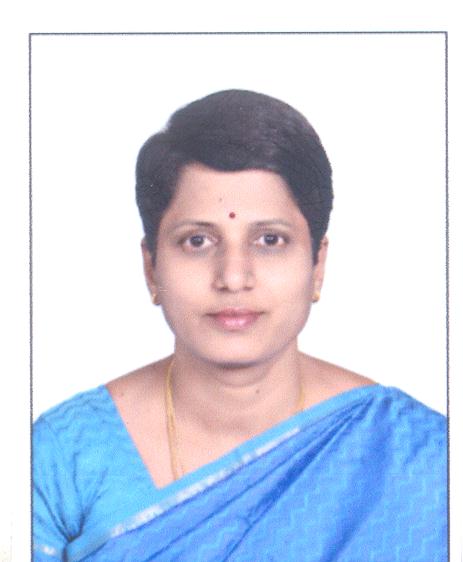}}{\textbf{Radhamani G.} is presently working as Professor and Director, School of  Information Technology and Science, Dr. G R Damodaran College of Science, affiliated to Bharathiar University, India. Formerly, she worked as Head, Department of IT, Ministry of Manpower, Sultanate of Oman. Prior to that she served as a Research Associate in IIT (India) and as a faculty in Department of Information Technology, Multimedia University, Malaysia. She received her M.Sc and M.Phil degrees from the P.S.G College of Technology, India, Ph.D degree from the Multimedia University, Malaysia. She had been invited to be Keynote Speaker and Chair for International conferences in India and abroad. She has published papers in International Journals and Conferences.}
		\end{biography}
		
		\begin{biography}{\includegraphics[width=75pt,height=80pt]{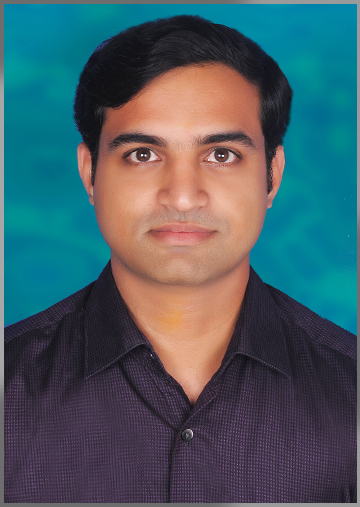}}{\textbf{Vinod P.} is Post Doc at Department of Mathematics, University of Padua, Italy. He holds his Ph.D in Computer Engineering from Malaviya National Institute of Technology, Jaipur, India. He has more than 70 research articles published in peer reviewed Journals and International Conferences. He is reviewer of number of security journals, and has also served as programme committee member in the International Conferences related to Computer and Information Security. His current research is involved in the development of malware scanner for mobile application using machine learning techniques. Vinod's area of interest is Adversarial Machine Learning, Malware Analysis, Context aware privacy persevering Data Mining, Ethical Hacking and Natural Language Processing.}
		\end{biography}
		
		\begin{biography}{\includegraphics[width=75pt,height=80pt]{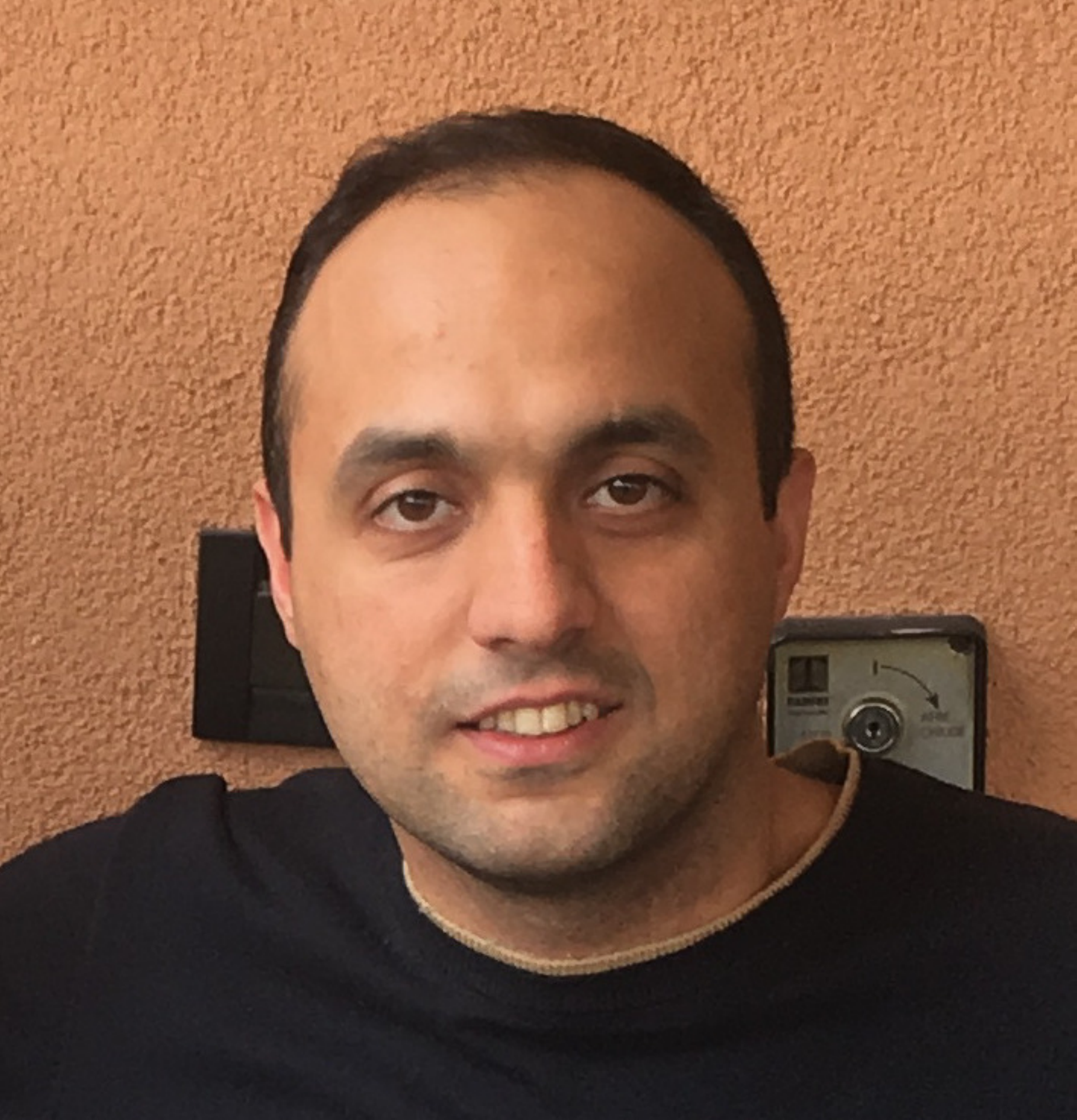}}{\textbf{Mohammad Shojafar} is an Intel Innovator and senior researcher in SPRITZ Security and Privacy Research Group at the University of Padua, Italy. He was CNIT Senior Researcher at the University of Rome Tor Vergata contributed on European H2020 ``SUPERFLUIDITY'' project. Also, he completed some Italian projects named ``SAMMClouds'', ``V-FoG'', ``PRIN15'' projects aim to address some of the open issues related to the Software as a Service (SaaS) and Infrastructure as a Service (IaaS) systems In Cloud and Fog computing which are supported by the University of Sapienza Rome and University of Modena and Reggio Emilia, Italy, respectively. He received the Ph.D. degree from Sapienza University of Rome, Rome, Italy, in 2016 with an ``Excellent'' degree. He received the MSc and BSc in QIAU and Iran University Science and Technology, Tehran, Iran in 2010 and 2006, respectively. He published over 90 refereed articles is prestigious venues such as IEEE TCC, IEEE TSC and IEEE TGCN. He was a programmer/analyzer at National Iranian Oil Company (NIOC) and Tidewater ltd in Iran from 2008-2013, respectively.}
		\end{biography}
		
		\begin{biography}{\includegraphics[width=75pt,height=86pt]{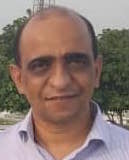}}{\textbf{Neeraj Kumar} is currently an Associate Professor in the Department of Computer Science and Engineering, Thapar University, Patiala (Pb.), India. He has published more than 200 technical research papers in leading journals and conferences from IEEE, Elsevier, Springer, John Wiley etc. Some of his research findings are published in top cited journals such as IEEE TIE, IEEE TDSC, IEEE TITS, IEEE TCE, IEEE Netw., IEEE Comm., IEEE WC, IEEE IoTJ, IEEE SJ, FGCS, JNCA, and ComCom. He has guided many research scholars leading to Ph.D. and M.E./M.Tech. His research is supported by fundings from Tata Consultancy Service, council of scientific and industrial research, and Department of Science \& Technology. He is a senior member of IEEE and committee member of different societies of ComSoc. He is in the editorial board member of IEEE Communication Magazine, Journal of Networks and Computer Applications, International Journal of Communication Systems, and Security and Privacy.}
		\end{biography}
		
		\begin{biography}{\includegraphics[width=75pt,height=86pt]{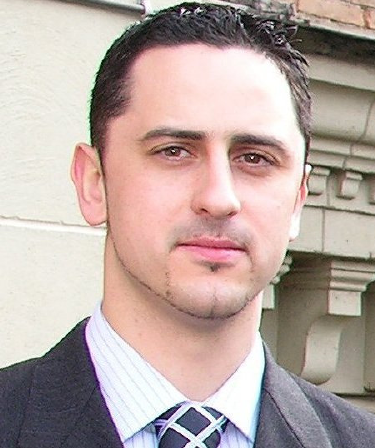}}{\textbf{Mauro Conti} is Full Professor at the University of Padua, Italy. He obtained his Ph.D. from Sapienza University of Rome, Italy, in 2009. After his Ph.D., he was a Post-Doc Researcher at Vrije Universiteit Amsterdam, The Netherlands. In 2011 he joined as Assistant Professor the University of Padua, where he became Associate Professor in 2015. In 2017, he obtained the national habilitation as Full Professor for Computer Science and Computer Engineering. He has been Visiting Researcher at GMU (2008, 2016), UCLA (2010), UCI (2012, 2013, 2014), TU Darmstadt (2013), UF (2015), and FIU (2015, 2016). He has been awarded with a Marie Curie Fellowship (2012) by the European Commission, and with a Fellowship by the German DAAD (2013). His main research interest is in the area of security and privacy. In this area, he published more than 200  papers in topmost international peer-reviewed journals and conference. He is Associate Editor for several journals, including IEEE Communications Surveys \& Tutorials and IEEE Transactions on Information Forensics and Security. He was Program Chair for TRUST 2015, ICISS 2016, WiSec 2017, and General Chair for SecureComm 2012 and ACM SACMAT 2013. He is Senior Member of the IEEE.}		
		\end{biography}
	\end{document}